\documentclass[12pt, draftclsnofoot, onecolumn]{IEEEtran}
 \usepackage{amsmath,amssymb}
 \usepackage{subfigure}
 \usepackage{graphicx,graphics,color,psfrag}
 \usepackage{cite,balance}
 \usepackage{caption}
 \allowdisplaybreaks
 \usepackage{accents}
 \usepackage{amsthm}
 \usepackage{bm}
 \usepackage[english]{babel}
 \usepackage{multirow}
 \usepackage{enumerate}
 \usepackage{cases}
 \usepackage{stfloats}
 \usepackage{dsfont}
 \usepackage{color,soul}
 \usepackage{amsfonts}
 \usepackage{cite,graphicx,amsmath,amssymb}
 \usepackage{subfigure}
 \usepackage{fancyhdr}
 \usepackage{hhline}
 \usepackage{graphicx,graphics}
 \usepackage{array,color}
 \usepackage{amsmath}
 \usepackage{booktabs}
 \usepackage{algorithmic,algorithm}
\newtheorem{theorem}{Theorem}

\newtheorem{lemma}{Lemma}

\newtheorem{proposition}{Proposition}

\newtheorem{remark}{\bf Remark}
\def\phi{\varphi}

\def\({\left(}
\def\){\right)}

\def\b0{{\mathbf{0}}}

\begin{document}
\setlength{\topskip}{-3pt}

\title{\huge Data Partition and Rate Control for Learning and Energy Efficient Edge Intelligence}
\author{Xiaoyang Li, \emph{Member, IEEE}, Shuai Wang, \emph{Member, IEEE}, Guangxu Zhu, \emph{Member, IEEE}, Ziqin Zhou, \emph{Student Member, IEEE}, Kaibin Huang, \emph{Fellow, IEEE}, and Yi Gong, \emph{Senior Member, IEEE}
\thanks{Xiaoyang Li, Shuai Wang, Ziqin Zhou, and Yi Gong are with the Department of Electrical and Electronic Engineering (EEE), Southern University of Science and Technology (SUSTech), Shenzhen, China. }
\thanks{Guangxu Zhu is with the Shenzhen Research Institute of Big Data, Shenzhen, China.}
\thanks{Kaibin Huang is with the Department of EEE, The University of Hong Kong, Hong Kong.} 
\thanks{Corresponding author: Yi Gong (gongy@sustech.edu.cn).} 
}
\maketitle

\vspace{-18mm}
\begin{abstract}
The rapid development of artificial intelligence together with the powerful computation capabilities of the advanced edge servers make it possible to deploy learning tasks at the wireless network edge, which is dubbed as \emph{edge intelligence} (EI). The communication bottleneck between the data resource and the server results in deteriorated learning performance as well as tremendous energy consumption. To tackle this challenge, we explore a new paradigm called \emph{learning-and-energy-efficient (LEE)} EI, which simultaneously maximizes the learning accuracies and energy efficiencies of multiple tasks via data partition and rate control. Mathematically, this results in a multi-objective optimization problem. Moreover, the continuous varying rates over the whole transmission duration introduce infinite variables. To solve this complex problem, we consider the case with infinite server buffer capacity and one-shot data arrival at sensor. First, the number of variables are reduced to a finite level by exploiting the optimality of constant-rate transmission in each epoch. Second, the optimal solution is found by applying \emph{stratified sequencing} or \emph{objectives merging}. By assuming higher priority of learning efficiency in stratified sequencing, the closed form of optimal data partition is derived by the Lagrange method, while the optimal rate control is proved to have the structure of \emph{directional water filling} (DWF), based on which a \emph{string-pulling} (SP) algorithm is proposed to obtain the numerical values. The DWF structure of rate control is also proved to be optimal in objectives merging via weighted summation. By exploiting the optimal rate changing properties, the SP algorithm is further extended to account for the cases with limited server buffer capacity or bursty data arrival at sensor. The performance of the proposed joint data partition and rate control design is examined by extensive experiments based on public datasets. 
\end{abstract}
\vspace{-5mm}


\section{Introduction}
Recent years have witnessed the revolutionary development of \emph{artificial intelligence} (AI) \cite{mitchell2013artificial}. Despite the powerful computation capability of cloud servers, completing AI tasks requires tremendous data samples for model training, which is expected to be provided by the ubiquitous \emph{Internet-of-Things} (IoT) devices \cite{zhou2019edge}. Nonetheless, sending vast amount of data from IoT devices to the cloud server causes a heavy communication burden in machine-type communication systems \cite{yu2020intelligent}. With the strengthened computation capability of edge servers (e.g., the network virtualization architecture standardized by 3GPP \cite{mao2017survey}), such a dilemma can be resolved by deploying AI tasks at the networks edge, which is known as \emph{edge intelligence} (EI) \cite{zhu2020toward}.

In contrast to the conventional designs for maximizing the throughput, the communication designs for EI aim at improving the learning performance \cite{wang2018edge}. The new objective has led to a set of new communication techniques \cite{zhu2020one,yang2020federated,amiri2020machine,zhang2021gradient,liu2020privacy,wang2020machine,ren2020accelerating,wen2020joint,chen2020joint,shi2020joint}. Realizing that the information required by learning tasks can be the statistics of data rather than the data itself, a series of researches advocate a new technique called over-the-air computation to support simultaneous access of multiple IoT devices and facilitate the data transmission for EI \cite{zhu2020one,yang2020federated,amiri2020machine,zhang2021gradient,liu2020privacy}. Another branch of works focuses on improving learning accuracy with the limited communication and/or computation resources \cite{wang2020machine,ren2020accelerating,chen2020joint,shi2020joint,wen2020joint}. To illustrate the impact of communication on learning accuracy, an empirical classification error model supported by learning theory was proposed in \cite{wang2020machine}, based on which a learning centric power allocation scheme was designed. To accelerate the learning process, the joint optimization of batch-size selection and resource allocation was investigated in \cite{ren2020accelerating}, while the joint design of computation load and bandwidth allocation was proposed in \cite{wen2020joint}. To improve the learning accuracy, the resource allocation and device scheduling were jointly optimized in \cite{chen2020joint} and \cite{shi2020joint}. 

Despite the extensive efforts on improving learning accuracy, energy consumption becomes non-negligible in both the communication and learning processes\cite{schwartz2020green}. To acquire enough energy, a sustainable learning scheme was proposed in \cite{guler2021sustainable} and \cite{guler2021energy} that leverages intermittent harvested energy from environments. Moreover, wireless power transfer was adopted in \cite{zeng2021wirelessly} to power the devices for data transmission and model training. Another vein to overcome the shortage of energy lies in the resource allocation designs for energy efficient EI \cite{sun2020energy,yang2020energy,mo2020energy,zeng2020energy}. By exploiting the data redundancy, an energy-aware analog transmission scheme for EI was designed in \cite{sun2020energy}. To minimize the energy consumption in EI, the computation and communication resource allocations were jointly optimized in \cite{yang2020energy} and \cite{mo2020energy}. In \cite{zeng2020energy}, the heterogenous CPU-GPU computation capabilities were exploited to improve the energy efficiency of EI.  It should be noted that the above works focus on myopic resource allocation for a single learning task, while a server may have multiple learning tasks to be executed sequentially in practice. The long-term transmission policy accounting for such scenario remains uncharted.  

The investigation of long-term transmission policy can be traced back to the earlier works on conventional communication networks. It was found in \cite{prabhakar2001energy} that the energy for passing given amount of data can be reduced by varying packet transmission time. Based on such finding, a long-term rate control scheme with \emph{string-pulling} (SP) structure was proposed to minimize the energy consumption under the quality of service constraints \cite{zafer2005calculus}, and was extended for the case with limited data buffer capacity \cite{zafer2009calculus}. Inspired by the SP structure, the subsequent work investigated long-term data transmission schemes for a variety of communication systems \cite{yang2011optimal,tutuncuoglu2012optimum,you2018exploiting}. In energy harvesting systems, the SP structure was proved to be optimal in power control for minimizing the transmission delay given the profiles of energy arrivals \cite{yang2011optimal} and battery capacity \cite{tutuncuoglu2012optimum}. In mobile edge computing systems, the optimal offloading data size over the whole computing duration for energy consumption minimization was determined based on the SP structure given the CPU-state information \cite{you2018exploiting}. Despite the rich literature on long-term data transmission policy in the literatures, there exists an additional data partition problem on top of the rate control to guarantee the learning accuracy for a EI system with multiple tasks. 

To incorporate the data partition in transmission design for EI, the current work investigates the effects of data sample size on the learning accuracy and energy efficiency. In particular, collecting more samples for model training can improve the learning performance especially when the server is at a shortage of data, while the data transmission brings extra energy consumption. Such effects lead to a tradeoff between the learning accuracy and energy consumption, where the former is characterized as a well-known classification error model in \cite{domhan2015speeding,johnson2018predicting,wang2020machine,liu2020edge,zhou2020learning} supported by the learning theory \cite{seung1992statistical}, and the latter is based on the Shannon’s theory. To achieve \emph{learning-and-energy efficient} (LEE) EI, the data partition and rate control for long-term transmission are jointly designed in this paper. The variables emerge in both the performance metrics of learning and energy, which causes nontrivial coupling and interdependence. By exploiting the optimal structure of data transmission as well as the optimization theory, the tractable designs are derived accounting for multiple scenarios, the performance of which are further examined by the extensive experiments on public datasets.


The main contributions of this work are summarized below.

\begin{itemize}
\item \emph{Joint Data Partition and Rate Control (JDPRC) Design:} Consider one-shot data arrival at sensor and infinite server data buffer capacity. To improve the learning accuracy as well as the energy efficiency, the data partition and rate control are jointly optimized, which results in a multi-objective optimization problem. Moreover, the continuous varying rates over the whole transmission duration introduce infinite variables. This complex problem is solved as follows. First, by exploiting the optimality of constant-rate transmission in each epoch, the number of variables are reduced to a finite level. Second, the optimal solution is found by applying \emph{stratified sequencing} or \emph{objectives merging}. By assuming higher priority of learning efficiency in stratified sequencing, the closed form of optimal data partition is derived by the Lagrange method, while the optimal rate control is proved to have the structure of \emph{directional water filling} (DWF), based on which a \emph{string-pulling} (SP) algorithm is proposed to obtain the numerical values. The DWF structure of rate control is also proved to be optimal in objectives merging via weighted summation.  
    
\item \emph{JDPRC with Limited Server Data Buffer Capacity:} The above solution approach is further extended to account for the case with limited server data buffer capacity. Geometrically, the optimal rate control policy involves finding a shortest path under the constraints of the required data size and data buffer capacity. By exploiting the optimal rate changing properties, a revised version of SP algorithm for rate control is proved to be optimal.
    
\item \emph{JDPRC with Bursty Data Arrival at Sensor:} The continuous data arriving curve at sensor adds infinite constraints on the rate control, which makes the optimization problem hard to be solved. For the tractability concern, the continuous data arriving curve is approximated by finite segments, based on which the SP algorithm is revised to obtain the transmission rates.
    
\item \emph{Validation by Testing on Public Datasets:} Extensive experimental results based on public datasets (including Scikit-learn, MNIST, Fashion MNIST, CIFAR-10, ModelNet-40) show that the proposed JDPRC is able to achieve higher accuracy as well as lower energy consumption than the schemes with the equal data partition and/or equal rate control.
\end{itemize}

The rest of this paper is organized as follows. The system model and problem formulation are described in Section II. The optimal design of data partition and rate control in the case with one-shot data arrival at sensor and infinite server data buffer capacity is presented in Section III, which is further extended to account for the cases with limited server data buffer capacity or bursty data arrival at sensor in Sections IV and V, respectively. The experimental results based on public datasets are presented in Section VI. The conclusions are drawn in Section VII.

\section{System Model and Problem Formulation}
\begin{figure}[t] 
\centering
\includegraphics[width=16cm]{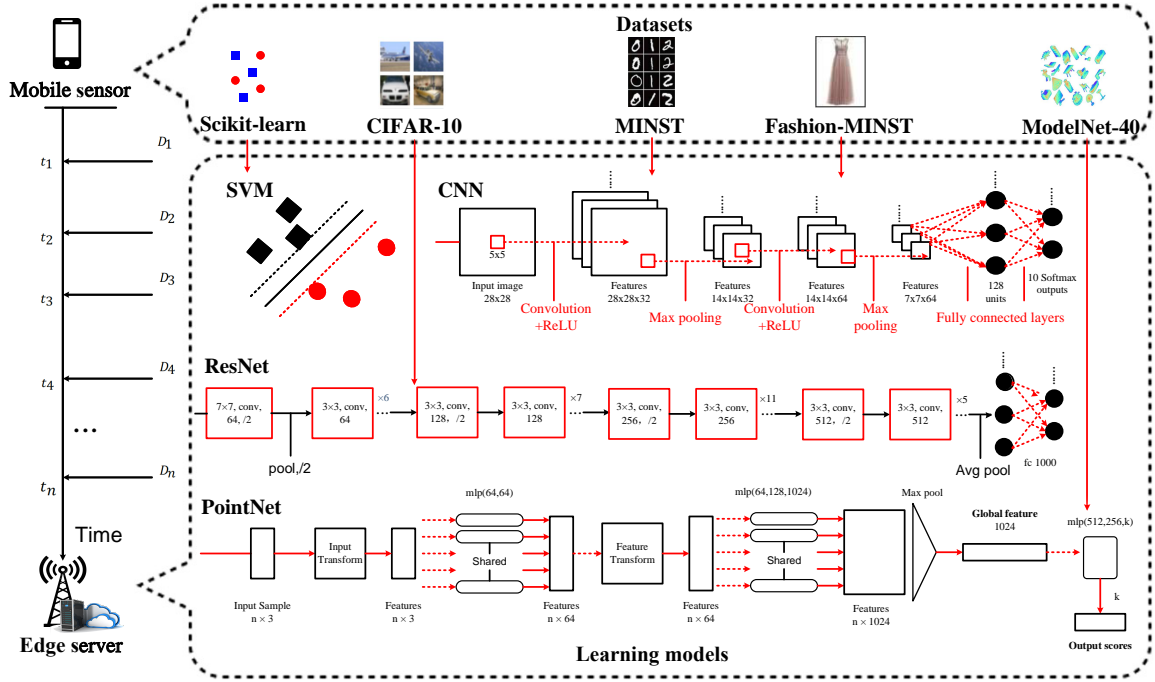}
\caption{EI system with multiple learning tasks.}
\label{FigSys}
\end{figure}

As depicted in Fig.~\ref{FigSys}, the EI system comprises one server and one sensor both equipped with single antenna. The server has $N$ model training tasks in a sequencing manner at instants $\{t_1,t_2,...t_N\}$. The number of data samples for training the $n$-th model at the server is
\begin{equation}\label{Eq:sample}
x_n = \left\lfloor\frac{D_n}{d_n}\right\rfloor + c_n \approx \frac{D_n}{d_n} + c_n,
\end{equation}
where $D_n$ represents the bits of data transmitted by the sensor for training the $n$-th model, $d_n$ represents the bits of data per sample, and $c_n$ denotes the previously stored samples at the server for the $n$-th model. The approximation is based on
$\lfloor x \rfloor \approx x$ when $x \gg 1$. The model training and data transmission processes are presented in the following sub-sections.

\subsection{Model Training Process}
The data samples are assumed to be independent and identically distributed. Therefore, the relationship between the learning error $e_n$ and the number of data samples $x_n$ can be depicted by the widely adopted inverse power law model in \cite{domhan2015speeding,johnson2018predicting,wang2020machine,liu2020edge,zhou2020learning}, which is supported by statistical mechanics of learning \cite{seung1992statistical} and expressed as
\begin{equation}\label{Eq:error}
e_n(x_n, a_n, b_n) = a_n x_n^{-b_n},
\end{equation}
where $a_n > 0$ and $b_n > 0$ are learning hyper-parameters. To train the $n$-th model with error no larger than $e_n$, at least $D_n$ bits of data need to be passed from the sensor to the server before $t_n$. The time interval between any two successive task instants is regarded as an epoch with the length $T_n = t_n - t_{n-1}$ for $n = 1,...N$ and $t_0 = 0$. The data is assumed to be already collected by and stored in the sensor at the beginning with amount $D$, which is also known as \emph{one-shot} data arrival at $t_0$. A more practical scenario w.r.t. \emph{bursty} data arrival is further analyzed in the subsequent sections, where the data sensing and transmission are simultaneously processed at the sensor. 

\subsection{Data Transmission Process}
To satisfy the requirements of model training, the sensor transmits data continuously and its rate can be varied via power control. Specifically, the sensor can choose transmission rate $r(t)$ at any instant $t$. The energy consumption for transmitting the required data by all $N$ learning tasks can be expressed as 
\begin{equation}\label{Eq:energy}
E =  \int_{t=0}^{t_N} \left(e^{r(t)/B}-1\right)\frac{\sigma^2}{h} dt,
\end{equation}
where $B$, $\sigma^2$, and $h$ denotes the bandwidth, noise, and channel power gain, respectively. To guarantee the execution of each task, the amount of transmitted data should be no less than the required one, i.e., 
\begin{equation}\label{Eq:lower}
\int_{t=0}^{t_j} r(t) dt \geq \sum_{n=1}^j D_n, j = 1,...N.~\text{(data transmission constraints)}
\end{equation}
If the size of data that can be transmitted in the $n$-th epoch is larger than the size of required data by the $n$-th task, such epoch can be used to transmit the data for subsequent tasks. The extra received data at the server can be stored in a data buffer. To gain the insights, the capacity of data buffer is assumed to be infinite and the channel condition is assumed to be fixed throughout the training duration. The analysis is further extended to the case with limited data buffer capacity. 

\subsection{Problem Formulation}
To achieve the energy efficient data transmission for high accuracy model training, both the learning errors and the energy consumption need to be minimized by optimizing data partition and rate control. Therefore, the performance metrics can be expressed as
\begin{equation}\label{Eq:objective}
\bold{f}(\{D_n\},r(t)) = [\{f_1(D_n)\},f_2(r(t))],
\end{equation}
where $f_1(D_n) = a_n \left(\frac{D_n}{d_n} + c_n\right)^{-b_n}$ and $f_2(r(t)) = \int_{t=0}^{t_N} \left(e^{r(t)/B}-1\right)\frac{\sigma^2}{h} dt$ denotes the learning errors and energy consumption, respectively. The corresponding optimization problem can be formulated as
\begin{subequations}
\begin{align}
\min_{\{D_n \geq 0\},\{r(t) \geq 0\}}  ~
& \bold{f}(\{D_n\},r(t)) \label{Eq:P1a}\\
\text{(P1)} \qquad \text{s.t.} \qquad
& \int_{t=0}^{t_j} r(t) dt \geq \sum_{n=1}^j D_n, j = 1,...N,\label{Eq:P1b}\\
& \sum_{n=1}^N D_n \leq D. \label{Eq:P1c}
\end{align}
\end{subequations}
Note that $\int_{t=0}^{t_N} r(t) dt = \sum_{n=1}^N D_n$ must hold, otherwise one can always decrease some $r(t)$ without conflicting any constraints, and thus reduce the energy consumption. The constraint in Eq.~\eqref{Eq:P1b} guarantees the transmission of required amount of data. The constraint in Eq.~\eqref{Eq:P1c} gives the total data budget. As there are multiple objectives in this problem, there exists a set of optimums at which improvingd the performance w.r.t. one objective will sacrifice that of others.  Moreover, since $r(t)$ is continuous in $[0,t_N]$, there are infinite optimization variables in problem (P1). To deal with such problem, a series of properties are exploited in the subsequent section.

\section{Optimal Data Partition and Rate Control}
In this section, the original problem is first simplified without loss of optimality by converting the continuous variables $\{r(t)\}$ into discrete variables $\{r_n\}$ based on the optimal structure of rate control. Next, to tackle the challenge of multiple objectives, two methods namely \emph{stratified sequencing} and \emph{objectives merging} are proposed. The former is to optimize the data partition and rate control in a sequential manner by assuming that the learning accuracy has high priority, while the later is based on the weighted summation of multiple objectives.


\subsection{Optimal Structure of Rate Control}
As for the objective of energy minimization, since the constant-rate transmission within each epoch is energy-efficient \cite{zafer2009calculus}, the optimal structure of rate control is given in the following lemma as proved in Appendix~\ref{App:Constant}.
\begin{lemma}[Optimality of Constant Transmission Rate]\label{Lemma:Constant}\emph{For any data partition $\{D_n\}$, a constant-rate transmission is optimal in a time interval $[t_a,t_b)$ if it satisfies the data transmission constraints in that interval.
}
\end{lemma}

Based on Lemma~\ref{Lemma:Constant}, the optimal data transmission rate $r(t)$ in the $n$-th epoch should be a constant denoted as $r_n$, and thus the problem (P1) can be simplified as:
\begin{subequations}
\begin{align}
\min_{\{D_n \geq 0\},\{r_n \geq 0\}}  ~
& \bold{f}(\{D_n\},\{r_n\}) \label{Eq:P2a}\\
\text{(P2)} \qquad \text{s.t.} \qquad
& \sum_{n=1}^j r_n T_n \geq \sum_{n=1}^j D_n, j = 1,...N,\label{Eq:P2b}\\
& \sum_{n=1}^N D_n \leq D. \label{Eq:P2c}
\end{align}
\end{subequations}

\subsection{Solution Approach based on Stratified Sequencing}
Suppose that the training error minimization has higher priority than energy consumption minimization, a \emph{stratified sequencing} method is proposed in this sub-section. The original problem (P2) can be divided into two sub-problems, where the first one aims at minimizing the weighted summation of classification errors via data partition, i.e.,
\begin{subequations}
\begin{align}
\min_{\{D_n \geq 0\}}  ~
& \sum_{n=1}^N \beta_n a_n \left(\frac{D_n}{d_n} + c_n\right)^{-b_n} \label{Eq:P2a1}\\
\text{(P2a)} \qquad \text{s.t.} \quad
& \sum_{n=1}^N D_n \leq D. \label{Eq:P2a2}
\end{align}
\end{subequations}
It can be easily observed that (P2a) is a convex problem with convex objective function and linear constraints. The Lagrange function of this problem can be expressed as
\begin{equation}\label{Eq:LagP2a}
L = \sum_{n=1}^N \beta_n a_n \left(\frac{D_n}{d_n} + c_n\right)^{-b_n} + \lambda(\sum_{n=1}^N D_n - D).
\end{equation}
Then, applying \emph{Karush-Kuhn-Tucker} (KKT) conditions leads to the following necessary and sufficient conditions:
\begin{subequations}
\begin{align}
& \frac{\partial L}{\partial D_n} = \lambda - \beta_n \frac{a_n b_n}{d_n} \left(\frac{D_n}{d_n} + c_n\right)^{-b_n-1} = 0, \label{Eq:KKTa1}\\
& \lambda(\sum_{n=1}^N D_n - D) = 0. \label{Eq:KKTa2}
\end{align}
\end{subequations}
Combining these conditions yields the optimal data partition given as 
\begin{equation}\label{Eq:OptData}
D_n^* = d_n\left(\frac{\beta_n a_n b_n}{\lambda^* d_n}\right)^{\frac{1}{b_n+1}} - c_n d_n.
\end{equation}
It can be observed that $\lambda \neq 0$ (otherwise $D_n^*$ is infinite large), and thus $\sum_{n=1}^N D_n = D$ must hold:
\begin{equation}\label{Eq:OptLambda}
\sum_{n=1}^N \left[d_n\left(\frac{\beta_n a_n b_n}{\lambda^* d_n}\right)^{\frac{1}{b_n+1}} - c_n d_n\right] = D.
\end{equation}
Since the left hand side monotonically decreases with the increasing $\lambda$, the optimal $\lambda^*$ can be derived by the bisection search as illustrated in Algorithm~\ref{Al:Bisection}. 

\begin{algorithm}[tt]
\renewcommand{\algorithmicrequire}{\textbf{Input:}}
\renewcommand{\algorithmicensure}{\textbf{Output:}}
\caption{Bisection Search for Optimal Data Partition.}
\label{Al:Bisection}
\begin{algorithmic}[1]
\REQUIRE tuning parameters $\{a_n\}$ and $\{b_n\}$, error weights $\{\beta_n\}$, size of data per sample $\{d_n\}$, and stored samples $\{c_n\}$.
\ENSURE the optimal data partition $\{D_n^*\}$.
\STATE Initialize $\lambda_{\ell}=10^{-5}$, $\lambda_{\ell}=10^5$, and $D(\lambda) = \sum_{n=1}^N \left[d_n\left(\frac{\beta_n a_n b_n}{\lambda d_n}\right)^{\frac{1}{b_n+1}} - c_n d_n\right]$.
\STATE Calculate $D(\lambda_{\ell})$ and $D(\lambda_h)$.
\STATE \textbf{While} $D(\lambda_h) - D(\lambda_{\ell}) > 10^{-10}$
\STATE \qquad Let $\lambda^*=(\lambda_{\ell}+\lambda_h)/2$, calculate $D(\lambda^*)$.
\STATE \qquad \textbf{If} $D(\lambda^*) < D$
\STATE \qquad \qquad Update $\lambda_h = \lambda^*$.
\STATE \qquad \textbf{Else}
\STATE \qquad \qquad Update $\lambda_{\ell} = \lambda^*$.
\STATE \qquad \textbf{End if}
\STATE \textbf{End while}
\STATE Calculate $D_n^* = d_n\left(\frac{\beta_n a_n b_n}{\lambda^* d_n}\right)^{\frac{1}{b_n+1}} - c_n d_n$.
\STATE \textbf{Return} the optimal data partition $\{D_n^*\}$.
\end{algorithmic}
\end{algorithm}
  
After determining the optimal $\{D_n^*\}$ by solving problem (P2a), the second sub-problem aims at energy consumption minimization via rate control:
\begin{subequations}
\begin{align}
\min_{\{r_n \geq 0\}}  ~
& \sum_{n=1}^{N} \left(e^{r_n/B}-1\right)\frac{\sigma^2 T_n}{h} \label{Eq:P2b1}\\
\text{(P2b)} \qquad \text{s.t.} \quad
& \sum_{n=1}^j r_n T_n \geq \sum_{n=1}^j D_n, j = 1,...N. \label{Eq:P2b2}
\end{align}
\end{subequations}
If can be observed that problem (P2b) is convex with convex objective and linear constraints. The Lagrange function of this problem can be expressed as
\begin{equation}\label{Eq:LagP2b}
L = \sum_{n=1}^{N} \left(e^{r_n/B}-1\right)\frac{\sigma^2 T_n}{h} + \sum_{j=1}^N \mu_j (\sum_{n=1}^j D_n - \sum_{n=1}^j r_n T_n).
\end{equation}
Then, applying KKT conditions leads to the following necessary and sufficient conditions:
\begin{subequations}
\begin{align}
& \frac{\partial L}{\partial r_n} = e^{r_n/B} \frac{\sigma^2 T_n}{hB} - \sum_{j=n}^N \mu_j T_n = 0, \label{Eq:KKTb1}\\
& \mu_j (\sum_{n=1}^j D_n - \sum_{n=1}^j r_n T_n) = 0,~j=1,...,N. \label{Eq:KKTb2}
\end{align}
\end{subequations}
Combining these conditions yields the optimal transmission rates given as 
\begin{equation}\label{Eq:OptRate}
r_n^* = B\left[\ln \frac{hB\sum_{j=n}^N \mu_j^*}{\sigma^2}\right]^+.
\end{equation}
It can be observed from Eq.~\eqref{Eq:OptRate} that the optimal rates depend on the Lagrange multipliers $\{\mu_j\geq 0\}$. To derive the specific values, the property of optimal rate control is exploited in the following lemma as proved in Appendix \ref{App:RateNon}.
\begin{lemma}[Optimality of Non-increasing Transmission Rates]\label{Lemma:RateNon}\emph{For any epoch $n$, the optimal transmission rate $r_n^*$ is monotonically decreasing, i.e., $r_{n+1}^* \leq r_n^*$. Moreover, if the data buffer is non-empty at the instant $t_j$, i.e., $\sum_{n=1}^j r_n T_n > \sum_{n=1}^j D_n$, the optimal transmission rate must obey that $r_{j+1}^* = r_j^*$.
}
\end{lemma}

\begin{figure}[t]
\centering
\includegraphics[scale=0.6]{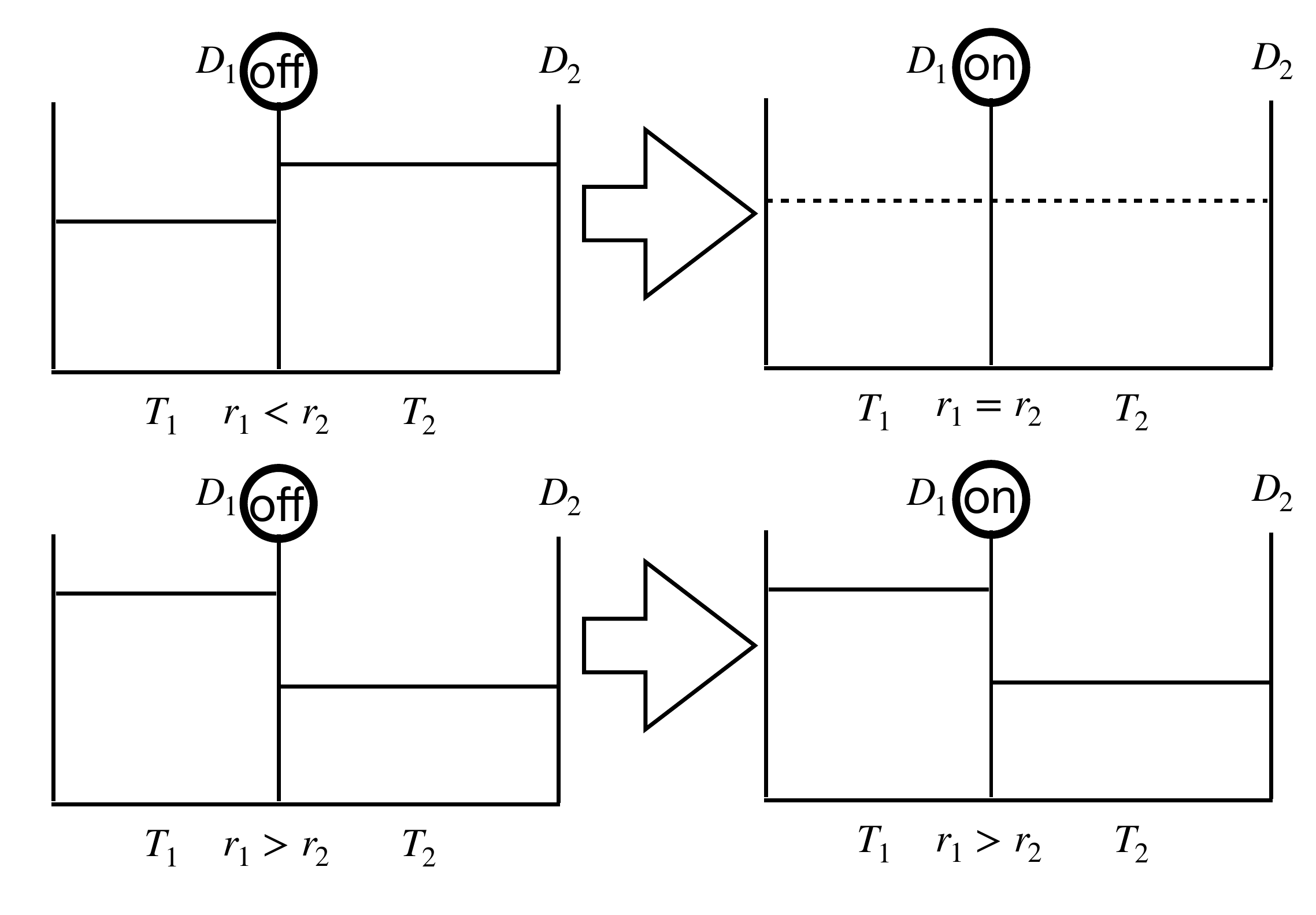}
\caption{Directional water-filling scheme for rate control.}
\label{FigDirectional}
\end{figure}

Such property can be referred to the \emph{directional water-filling} scheme in energy harvesting investigation \cite{ozel2011transmission}. As shown in the top figure of Fig.~\ref{FigDirectional}, if $r_1 < r_2$, then part of data for subsequent learning tasks can be passed in the former epoch so that the levels can be equalized. However, if $r_1 > r_2$, no data can flow from left to right since the data received in later epochs cannot be used for training the previous models. Therefore, as shown in the bottom figure of Fig.~\ref{FigDirectional}, the water levels are not equalized. 

\begin{figure}[t]
\centering
\includegraphics[scale=0.6]{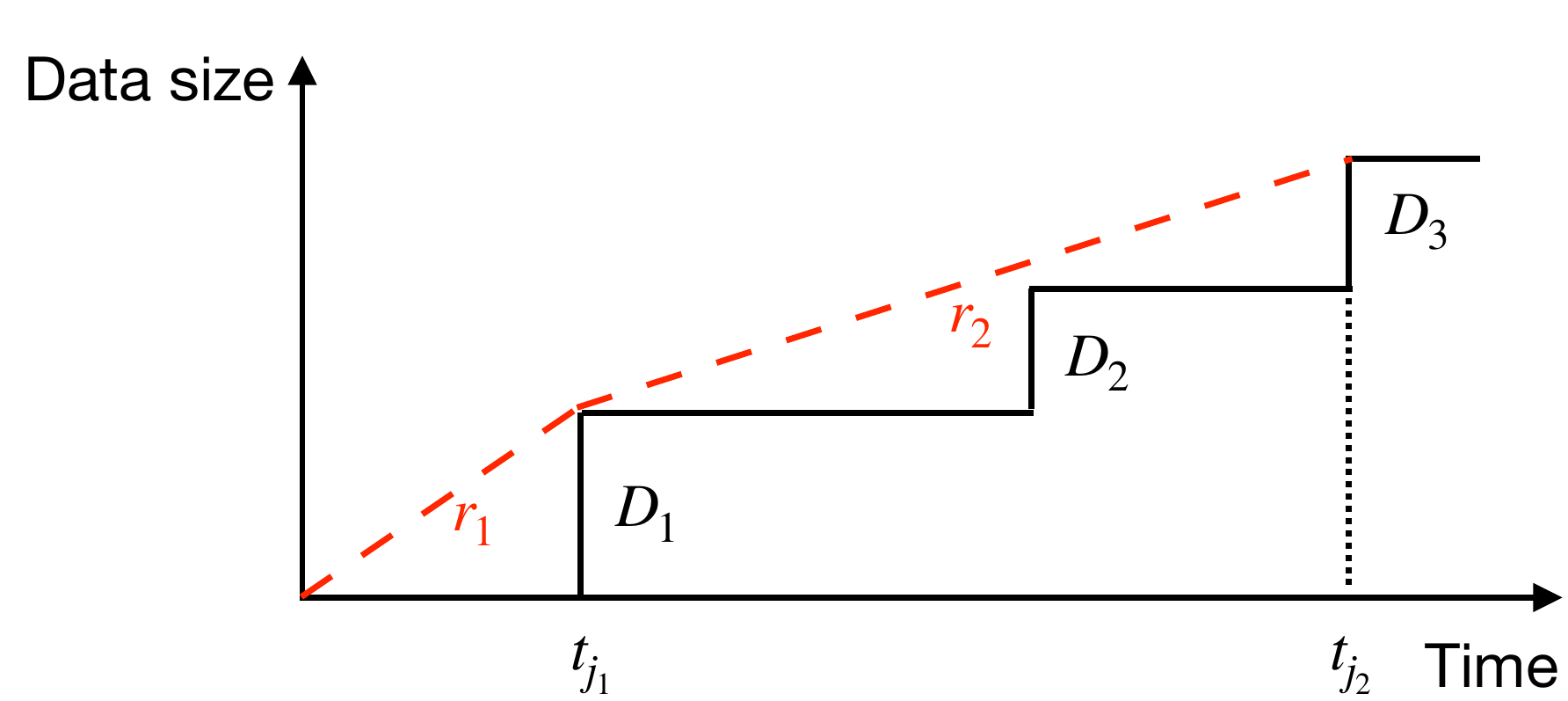}
\caption{An interpretation of rate control satisfying Lemmas~\ref{Lemma:Constant} and \ref{Lemma:RateNon}.}
\label{FigRate}
\end{figure}

Based on Lemmas~\ref{Lemma:Constant} and \ref{Lemma:RateNon}, one can characterize the optimal rate control in the following way. Given task instants $\{t_n\}$ and learning hyper-parameters $\{a_n,b_n\}$, the required size of data is plotted as a function of $t$, which is a staircase curve in Fig.~\ref{FigRate}. The total amount of transmitted data up to time $t$ can also be represented as a continuous curve, depicted by the dashed line in Fig.~\ref{FigRate}. In order to satisfy the data transmission constraints, the curve of transmitted data size must lie above the curve of required data size at all times. Based on Lemma~\ref{Lemma:Constant}, the optimal curve of transmitted data size must be linear in each epoch, and the slope of the segment corresponds to the transmission rate. Lemma~\ref{Lemma:RateNon} implies that whenever the slope changes, the curve of transmitted data size must touch the curve of required data size at that instant. Therefore, the first linear segment of the curve of transmitted data size must be one of the lines connecting the origin and a corner point on the curve of required data size. Because of the monotonicity property of the rate given in Lemma~\ref{Lemma:RateNon}, among these lines, the one with the maximal slope should be picked. Otherwise, either the data transmission constraints or the monotonicity property given in Lemma~\ref{Lemma:RateNon} will be violated. Based on the above analysis, the structure of the optimal policy can be expressed in the following proposition as proved in Appendix~\ref{App:OptRate}. 

\begin{proposition}[Optimal Rate Control Policy]\label{Prop:OptRate}\emph{Consider a policy with transmission rate vector $\mathbf{r} = [r_1, r_2, ..., r_i,...]$ and corresponding duration vector $\mathbf{T} = [T_1, T_2, ..., T_i...]$. This policy is optimal if and only if it has the following structure:
\begin{subequations}
\begin{align}
& j_i = \arg\max_{j:j_{i-1}<j \leq N}\left\{\frac{\sum_{n=j_{i-1}+1}^{j}D_n}{t_j-t_{j_{i-1}}}\right\}, \label{Eq:Unlimited1} \\ 
& r_i = \left\{\frac{\sum_{n=j_{i-1}+1}^{j_i}D_n}{t_{j_i}-t_{j_{i-1}}}\right\}, \label{Eq:Unlimited2} \\
& T_i = t_{j_i}-t_{j_{i-1}}, \label{Eq:Unlimited3}
\end{align}
\end{subequations}
where $j_i$ is the index of instant when the rate $r_i$ is switched to $r_{i+1}$.
}
\end{proposition}

From the results, one can conclude that given the required data amounts $\{D_n\}$ and starting instants $\{t_n\}$ of learning tasks, the optimal rate control policy is known via Proposition~\ref{Prop:OptRate}. In particular, the optimal rate control policy is the one that yields the tightest piecewise linear curve of transmitted data size that lies above the curve of required data size at all times and touches the curve of required data size at $\{t_{j_i}\}$, which is known as \emph{string-pulling}. The procedure to obtain the optimal rate control policy is summarized in Algorithm~\ref{Al:OptUnlimited}. 
\begin{algorithm}[tt]
\renewcommand{\algorithmicrequire}{\textbf{Input:}}
\renewcommand{\algorithmicensure}{\textbf{Output:}}
\caption{String Pulling for Optimal Rate Control.}
\label{Al:OptUnlimited}
\begin{algorithmic}[1]
\REQUIRE required data amounts $\{D_n\}$ at instants $\{t_n\}$ of $N$ learning tasks.
\ENSURE the optimal transmission rates $\{r_i^*\}$ and durations $\{T_i^*\}$.
\STATE Initialize $j_0 = 0$, $i = 0$.
\STATE \textbf{while} $j_i < N$
\STATE \qquad Update $i = i+1$.
\STATE \qquad Calculate $j_i = \arg\max_{j:j_{i-1}<j\leq N}\left\{\frac{\sum_{n=j_{i-1}+1}^{j}D_n}{t_j-t_{j_{i-1}}}\right\}$. 
\STATE \qquad Calculate $r_i^*= \left\{\frac{\sum_{n=j_{i-1}+1}^{j_i}D_n}{t_{j_i}-t_{j_{i-1}}}\right\}$.
\STATE \qquad Calculate $T_i^* = t_{j_i}-t_{j_{i-1}}$.
\STATE \textbf{End while}
\STATE \textbf{Return} the optimal transmission rates $\{r_i^*\}$ and durations $\{T_i^*\}$.
\end{algorithmic}
\end{algorithm}

One could observe from Fig.~\ref{FigRate} that the $n$-th transmission rate constraint is inactive if $D_n/T_n \leq D/t_N$. For the case with $D_n/T_n \leq D/t_N$ for all $n$, problem (P2b) is degraded as a classical BT-problem with $B$ amount of data to be transmitted by deadline $T$. According to \cite{zafer2005calculus}, the corresponding optimal transmission rate should be a constant determined by $r^* = D/t_N$. It should be noted that the optimal data partition and rate control derived by stratified sequencing is a point on the Pareto boundary with the assumption that training error minimization has higher priority than energy consumption minimization. To characterize other points on the Pareto boundary, the objectives merging method is applied in the next sub-section.

\subsection{Solution Approach based on Objectives Merging}
As an alternative approach for solving the multi-objective problem, the weighted summation is applied to merge the multiple objectives into one. The weights of energy consumption and training errors are denoted by $\alpha$ and $\{\beta_n\}$ respectively, which satisfy that $\alpha + \sum_{n=1}^N \beta_n = 1$. The corresponding problem can be expressed as
\begin{subequations}
\begin{align}
\min_{\{D_n \geq 0\},\{r_n \geq 0\}}  ~
& \alpha \sum_{n=1}^{N} \left(e^{r_n/B}-1\right)\frac{\sigma^2 T_n}{h} + \sum_{n=1}^N \beta_n a_n \left(\frac{D_n}{d_n} + c_n\right)^{-b_n} \label{Eq:P4a}\\
\text{(P3)} \qquad \text{s.t.} \qquad
& \sum_{n=1}^j r_n T_n \geq \sum_{n=1}^j D_n, j = 1,...N,\label{Eq:P4b}\\
& \sum_{n=1}^N D_n \leq D. \label{Eq:P4c}
\end{align}
\end{subequations}
It can be easily proved that problem (P3) is a convex problem with convex objective and linear constraints. The corresponding Lagrangian can be expressed as
\begin{equation}\label{Eq:Lagrangian}
L \!=\! \alpha \sum_{n=1}^{N} \left(e^{r_n/B}-1\right)\frac{\sigma^2 T_n}{h} \!+\! \sum_{n=1}^N \beta_n a_n \left(\frac{D_n}{d_n} \!+\! c_n\right)^{-b_n} \!+\! \sum_{j=1}^N \mu_j (\sum_{n=1}^j D_n \!-\! \sum_{n=1}^j r_n T_n) \!+\! \lambda(\sum_{n=1}^N D_n \!-\! D)
\end{equation}
Then, applying KKT conditions leads to the following necessary and sufficient conditions:
\begin{subequations}
\begin{align}
& \frac{\partial L}{\partial r_n} = \alpha e^{r_n/B} \frac{\sigma^2 T_n}{hB} - \sum_{j=n}^N \mu_j T_n = 0, \label{Eq:KKT1}\\
& \frac{\partial L}{\partial D_n} = \lambda + \sum_{j=n}^N \mu_j - \beta_n \frac{a_n b_n}{d_n} \left(\frac{D_n}{d_n} + c_n\right)^{-b_n-1} = 0, \label{Eq:KKT2}\\
& \mu_j (\sum_{n=1}^j D_n - \sum_{n=1}^j r_n T_n) = 0, j=1,...,N, \label{Eq:KKT3}\\
& \lambda(\sum_{n=1}^N D_n - D) = 0. \label{Eq:KKT4}
\end{align}
\end{subequations}
Combining these conditions yields the optimal data partition and rate control policy given in the following proposition.
\begin{proposition}[Optimal Data Partition and Rate Control Policy]\label{Pro:OptDPRC}\emph{The optimal data partition and rate control policy for solving problem (P4) are given as
\begin{subequations}
\begin{align}
D_n^* & = d_n\left(\frac{\beta_n a_n b_n}{\left(\lambda^* + \sum_{j=n}^N \mu_j^*\right)d_n}\right)^{\frac{1}{b_n+1}} - c_n d_n, \label{Eq:data} \\
r_n^* & = B\left[\ln \frac{hB\sum_{j=n}^N \mu_j^*}{\alpha\sigma^2}\right]^+.\label{Eq:rate}
\end{align}
\end{subequations}
}
\end{proposition}

\begin{remark}[Optimal Structure in Joint Optimization after Objectives Merging]\emph{It can be observed from Eq.~\eqref{Eq:rate} that the optimality of non-decreasing rates given by Lemma~\ref{Lemma:RateNon} still holds for the joint optimization of data partition and rate control after objectives merging. The specific values of $D_n^*$ and $r_n^*$ can be obtained by the CVX Toolbox \cite{grant2009cvx} for convex programming.
}
\end{remark}

\section{Data Partition and Rate Control for Limited Buffer Capacity Case}
Inspired by the policy in the previous section, the data partition and rate control for limited data buffer capacity case are investigated in this section. Constrained by the maximum capacity $D_{\max}$ of the buffer, the size of stored data should obey the buffer constraints:
\begin{equation}\label{Eq:upper}
\sum_{n=1}^j r_n T_n - \sum_{n=0}^{j-1} D_n \leq D_{\max}, j = 1,...N,~\text{(data buffer constraints)}
\end{equation}
where $D_0 = 0$ represents that no data is required at instant $t_0$. After incorporating the data buffer constraints, the optimization problem can be formulated as
\begin{subequations}
\begin{align}
\min_{\{D_n \geq 0\},\{r_n \geq 0\}}  ~
& \bold{f}(\{D_n\},\{r_n\}) \label{Eq:P4a}\\
\text{(P4)} \qquad \text{s.t.} \qquad
& \sum_{n=1}^j r_n T_n \geq \sum_{n=1}^j D_n, j = 1,...N,\label{Eq:P4b}\\
& \sum_{n=1}^j r_n T_n - \sum_{n=0}^{j-1} D_n \leq D_{\max}, j = 1,...N, \label{Eq:P4c}\\
& D_n \leq D_{\max}, n =1,.., N, \label{Eq:P4d}\\
& \sum_{n=1}^N D_n \leq D. \label{Eq:P4c}
\end{align}
\end{subequations}
Note that the constraints in Eq.~\eqref{Eq:P4d} guarantees that the required amount of data in each epoch cannot exceed the data buffer capacity. Such problem can also be solved by the stratified sequencing or objectives merging as elaborated below.

\subsection{Stratified Sequencing for the Case of Limited Buffer Capacity}
Similar to the solving approach for problem (P2), problem (P4) can also be divided into two sub-problems w.r.t. learning errors minimization and energy consumption minimization. The former sub-problem has the exact form of problem (P2a) plus the constraints in Eq.~\eqref{Eq:P4d} and can be solved by the CVX Toolbox, while solving the latter one requires exploiting the properties in the following lemmas proved in Appendices~\ref{App:Overflow}, \ref{App:Condition}, and \ref{App:Direction}.
\begin{lemma}[Sub-optimality of Data Buffer Overflow]\label{Lemma:Overflow}\emph{Any transmission rate policy yielding a data buffer overflow is strictly sub-optimal.
}
\end{lemma}

\begin{lemma}[Rate Changing Condition]\label{Lemma:Condition}\emph{In the optimal transmission rate policy, the transmission rate does not change unless the data buffer is either full or empty.
}
\end{lemma}

\begin{lemma}[Rate Changing Direction]\label{Lemma:Direction}\emph{For optimal data transmission design, the change in rate $r(t)$ cannot be negative (or positive) unless the data buffer is empty (or full) at this instant.
}
\end{lemma}



Denote the subsequence of $\{t_n\}$ at which the transmission rate changes as $\{t_{u_i}\}$, the transmission rate has to be the form of
\begin{equation}
r(t) = r_i,~\forall t \in [t_{u_{i-1}}, t_{u_i}).
\end{equation}

Note that once the specifics of the first interval $[0, t_{u_1}]$ is determined, the remaining of the problem can be considered as a separate energy minimization problem. That is, given the duration of this interval $t_{u_1}$, and the amount of energy consumption in this epoch $\left(e^{r_1/B}-1\right)\frac{\sigma^2 t_{u_1}}{h}$, it remains to solve for the optimal rate from a modified problem with task starting instants shifted by $t_{u_1}$ and a new initial data buffer state $r_1 t_{u_1} - \sum_{n=0}^{u_1} D_n$. This means that once the first slot of the optimal rate is identified, the remaining rates can be found recursively with the same algorithm using updated parameters. Therefore, we shall focus on determining the optimal rate in the initial epoch. The modified problem described above is known as the \emph{shifted} optimization problem.

We define two sets of rates $\{r_e[1],r_e[2]...\}$ and $\{r_f[1],r_f[2]...\}$, where $r_e[j]$ and $r_f[j]$ are the constant rates that would result in an empty data buffer at $t_j^+$ or a full data buffer at $t_j^-$ respectively if employed in $[0,t_j]$. We then define the set $\bold{r} = \{\bold{r}[1],\bold{r}[2]...\}$ with elements as the closed intervals $\bold{r}[j] = [r_e[j],r_f[j]]$ between corresponding elements of the two sets $\{r_e\}$ and $\{r_f\}$. This translates to a range of constant rates that would be feasible for the $j$-th epoch when the feasibility at previous epochs are disregarded. Therefore, we have
\begin{subequations}\label{Eq:interval}
\begin{align}
& r_e[j] = \frac{\sum_{n=0}^j D_n}{t_j},\\
& r_f[j] = \frac{\sum_{n=0}^{j-1} D_n + D_{\max}}{t_j},\\
& \bold{r}[j] = [r_e[j],r_f[j]] = \{r|r_e[j] \leq r \leq r_f[j]\},
\end{align}
\end{subequations}
for $j = 1,...,N$. Based on this definition of the feasible rate range, it can be deducted that for a constant rate transmission starting from $t = 0$ to the $u$-th task starting instant without violating data feasibility, its rate should be contained in the range $\bold{r}[j]$ for $j = 1,...,u$, i.e., the step should be feasible through all display it extends over. This yields an upper bound $u_b$ on the length of the first constant rate transmission, which can be calculated as
\begin{equation}\label{Eq:bound}
u_b = \max\left\{u|\bigcap_{j=1}^u\bold{r}[j]\neq \emptyset, j=1,2,...,N\right\}.
\end{equation}
As for the later displays, a feasible constant rate contained in all previous feasible rate ranges do not exist. Given the sets of $\{r_e[j]\}$ and $\{r_f[j]\}$, assume that a constant rate $r_1$ and duration $t_{u_1}$ is feasible. This transmission then satisfies $r_1 \in \bigcap_{j=1}^{u_1}\bold{r}[j]$ and cannot extend beyond $t_{u_b}$, as it is rendered infeasible at $t_{u_b+1}$ by one of the constraints.

\begin{figure}[t]
\centering
\includegraphics[scale=0.6]{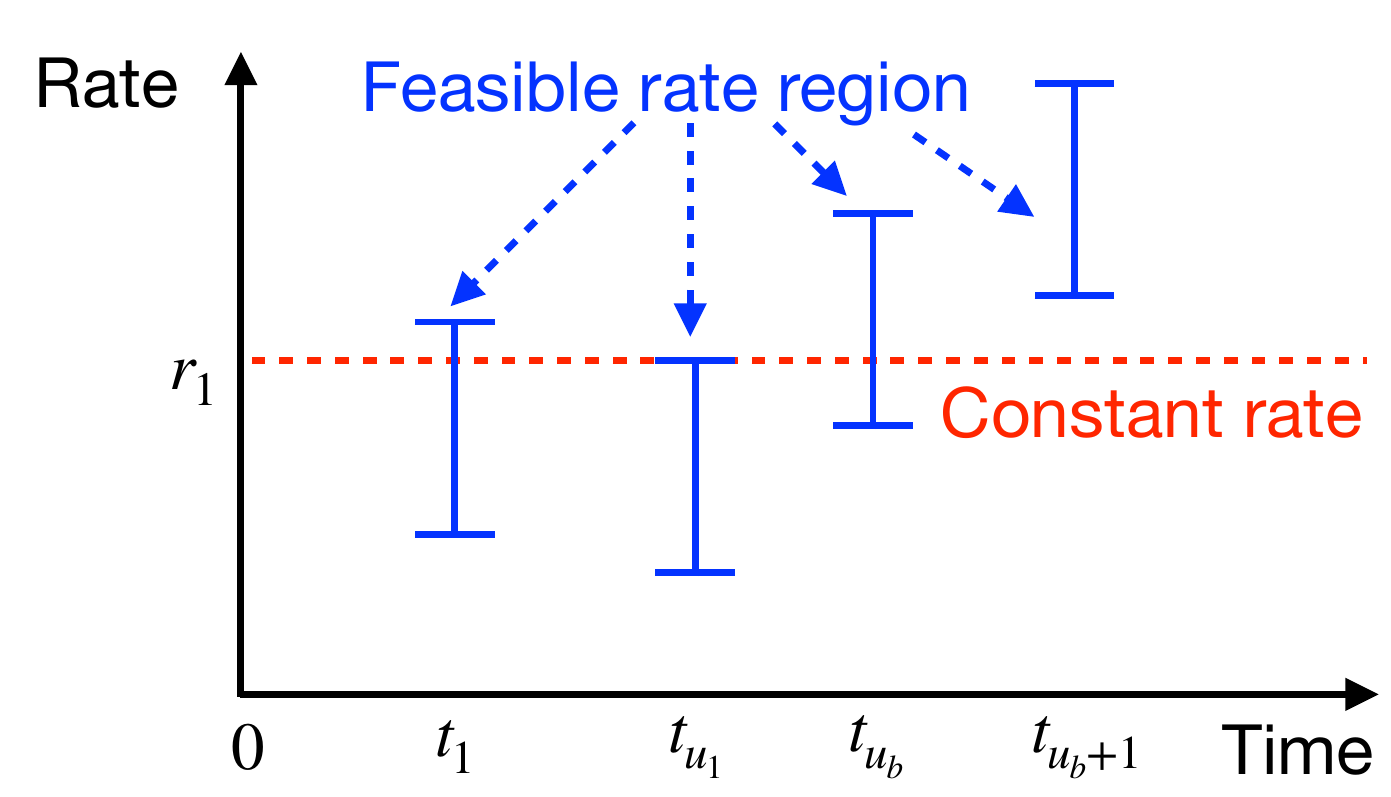}
\caption{The range of feasible rates.}
\label{Figr}
\end{figure}

As shown in Fig.~\ref{Figr}, a transmission with constant-rate $r_1$ either fails to satisfy the required data size or overflows the data buffer at $t_{u_b+1}$. The former case implies that the rate after $t_{u_1}$ needs to increase, and the latter implies that the rate needs to decrease. This can be verified by calculating updated values of $r_e[j]$ and $r_f[j]$ for a shifted problem after the first step of the policy is determined. By Lemma \ref{Lemma:Direction}, an increase or decrease in rate can occur only at a instant with the data buffer empty or full. Hence, the choice of $r_1$ in the optimal policy is restricted to $r_e[u_1]$ or $r_f[u_1]$ respectively for the two cases in consideration, where $u_1 = \max\left\{u|r_e[u] \in \bigcap_{j=1}^{u_b}\bold{r}[j]\right\}$ if $\bold{r}[u_b+1]$ falls below $\bigcap_{j=1}^{u_b}\bold{r}[j]$, otherwise $u_1 = \max\left\{u|r_f[u] \in \bigcap_{j=1}^{u_b}\bold{r}[j]\right\}$. Based on the above findings, Algorithm~\ref{Al:OptLimited} is designed for energy efficient rate control with the optimality given in the following theorem as proved in Appendix~\ref{App:OptLimited}.
\begin{theorem}[Optimal Rate Control for Limited Data Buffer Capacity Case]\label{Theorem:OptLimited}\emph{Algorithm~\ref{Al:OptLimited} yields the optimal transmission rate control.
}
\end{theorem}

\begin{algorithm}[tt]
\renewcommand{\algorithmicrequire}{\textbf{Input:}}
\renewcommand{\algorithmicensure}{\textbf{Output:}}
\caption{String Pulling for Rate Control in Limited Data Buffer Case.}
\label{Al:OptLimited}
\begin{algorithmic}[1]
\REQUIRE  required data amounts $\{D_n\}$ at instants $\{t_n\}$ of $N$ tasks, data buffer capacity $D_{\max}$. 
\ENSURE the optimal transmission rates $\{r_i^*\}$ and durations $\{T_i^*\}$.
\STATE Initialize $u_b = 0$, $u_1 = 0$, $i = 0$.
\STATE \textbf{while} $N > 0$
\STATE \qquad Update $i = i+1$.
\STATE \qquad Update $r_e[j]$, $r_f[j]$, and $\bold{r}[j]$ for $j = 1,...,N$ according to \eqref{Eq:interval}.
\STATE \qquad Update $u_b$ according to \eqref{Eq:bound}.
\STATE \qquad \textbf{if} $u_b = N$ 
\STATE \qquad \qquad Update $u_1 = \max\left\{u|r_e[u] \in \bigcap_{j=1}^{u_b}\bold{r}[j]\right\}$, $r_i^* = r_e[u_1]$, $T_i^* = t_{u_1}$.
\STATE \qquad \textbf{else} 
\STATE \qquad \qquad \textbf{if} $\bold{r}[u_b+1]$ falls below $\bigcap_{j=1}^{u_b}\bold{r}[j]$
\STATE \qquad \qquad \qquad Update $u_1 = \max\left\{u|r_e[u] \in \bigcap_{j=1}^{u_b}\bold{r}[j]\right\}$, $r_i^* = r_e[u_1]$, $T_i^* = t_{u_1}$.
\STATE \qquad \qquad \textbf{else}
\STATE \qquad \qquad \qquad Update $u_1 = \max\left\{u|r_f[u] \in \bigcap_{j=1}^{u_b}\bold{r}[j]\right\}$, $r_i^* = r_f[u_1]$, $T_i^* = t_{u_1}$.
\STATE \qquad \qquad \textbf{End if}
\STATE \qquad \textbf{End if}
\STATE \qquad Update $N = N - u_1$, $t_n = t_{n+u_1} - t_{u_1}$, $D' = r_i^* T_i^* - \sum_{n=0}^{u_1} D_n$.
\STATE \qquad Update $D_n = D_{n+u_1}$, $D_1 = D_1 - D'$.
\STATE \textbf{End while}
\STATE \textbf{Return} the optimal transmission rates $\{r_i^*\}$ and durations $\{T_i^*\}$.
\end{algorithmic}
\end{algorithm}

\begin{figure}[t]
\centering
\includegraphics[scale=0.6]{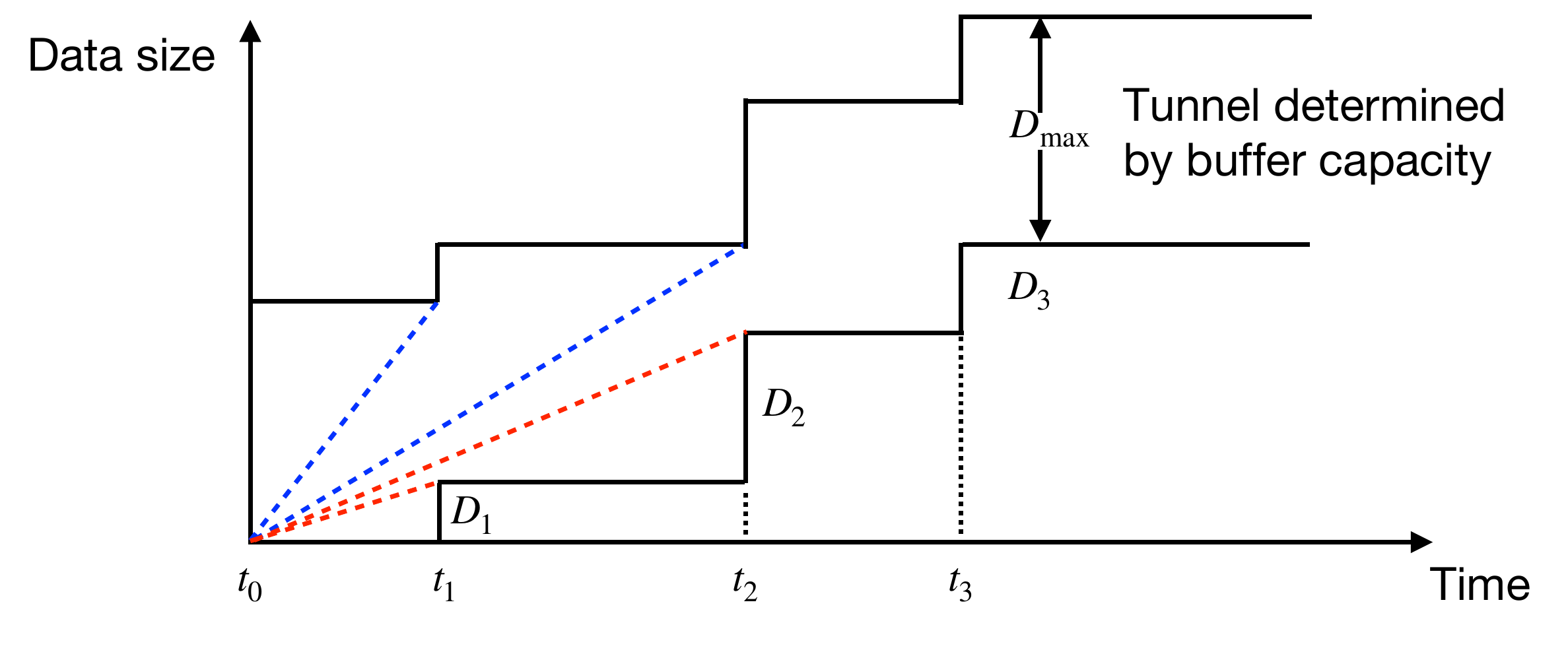}
\caption{The feasible data transmission tunnel with limited data buffer capacity.}
\label{FigTunnel}
\end{figure}

The \emph{feasible data transmission tunnel with limited data buffer capacity} is shown in Fig.~\ref{FigTunnel}. The lower solid boundary represents the required data size for learning tasks and the upper solid boundary is the lower boundary shifted up by an amount of $D_{\max}$. The cumulative data transmitted by Algorithm~\ref{Al:OptLimited} forms a continuous line, and must stay within this tunnel. A transmission rate that goes above the tunnel causes a data buffer overflow, while one that goes below the tunnel fails to transmit the required amount of data. The sets $\{r_e[j]\}$ and $\{r_f[j]\}$ correspond to the slopes of lines from the origin to each of the corner points in the tunnel as shown with dashed lines. The algorithm first determines the longest constant-rate transmission that stays within this tunnel, and then determines whether the furthest point on a wall that a line passing through the origin can reach is an upper bound or a lower bound. This is accomplished by comparing the first unreachable interval $\bold{r}[u_b+1]$ with the earlier ones. Finally, the algorithm selects the longest feasible constant-rate transmission that ends in one of the sets $\{r_e[j]\}$ and $\{r_f[j]\}$, allowing a change in transmission rate for the rest of the problem. If the optimal data partition satisfies that $D_n^*/T_n \leq D/t_N \leq (D_n^*+D_{\max})/T_n$ for all $n$, then optimal rate should be a constant determined by $r^* = D/t_N$.

\subsection{Objectives Merging for the Case of Limited Buffer Capacity}
By applying weighted summation for objectives merging, the problem can be formulated as
\begin{subequations}
\begin{align}
\min_{\{D_n \geq 0\},\{r_n \geq 0\}}  ~
& \alpha \sum_{n=1}^{N} \left(e^{r_n/B}-1\right)\frac{\sigma^2 T_n}{h} + \sum_{n=1}^N \beta_n a_n \left(\frac{D_n}{d_n} + c_n\right)^{-b_n} \label{Eq:P5a}\\
\text{(P5)} \qquad \text{s.t.} \qquad
& \sum_{n=1}^j r_n T_n \geq \sum_{n=1}^j D_n, j = 1,...N,\label{Eq:P5b}\\
& \sum_{n=1}^j r_n T_n - \sum_{n=0}^{j-1} D_n \leq D_{\max}, j = 1,...N, \label{Eq:P4c}\\
& D_n \leq D_{\max}, n = 1,...,N,\label{Eq:P5d}\\
& \sum_{n=1}^N D_n \leq D. \label{Eq:P5e}
\end{align}
\end{subequations}
It can be easily proved that problem (P5) is a convex problem with convex objective and linear constraints. Given the specified weights, the problem can be solved by the CVX Toolbox.


\section{Data Partition and Rate Control for Bursty Data Arrival Case}
\begin{figure}[t]
\centering
\includegraphics[scale=0.6]{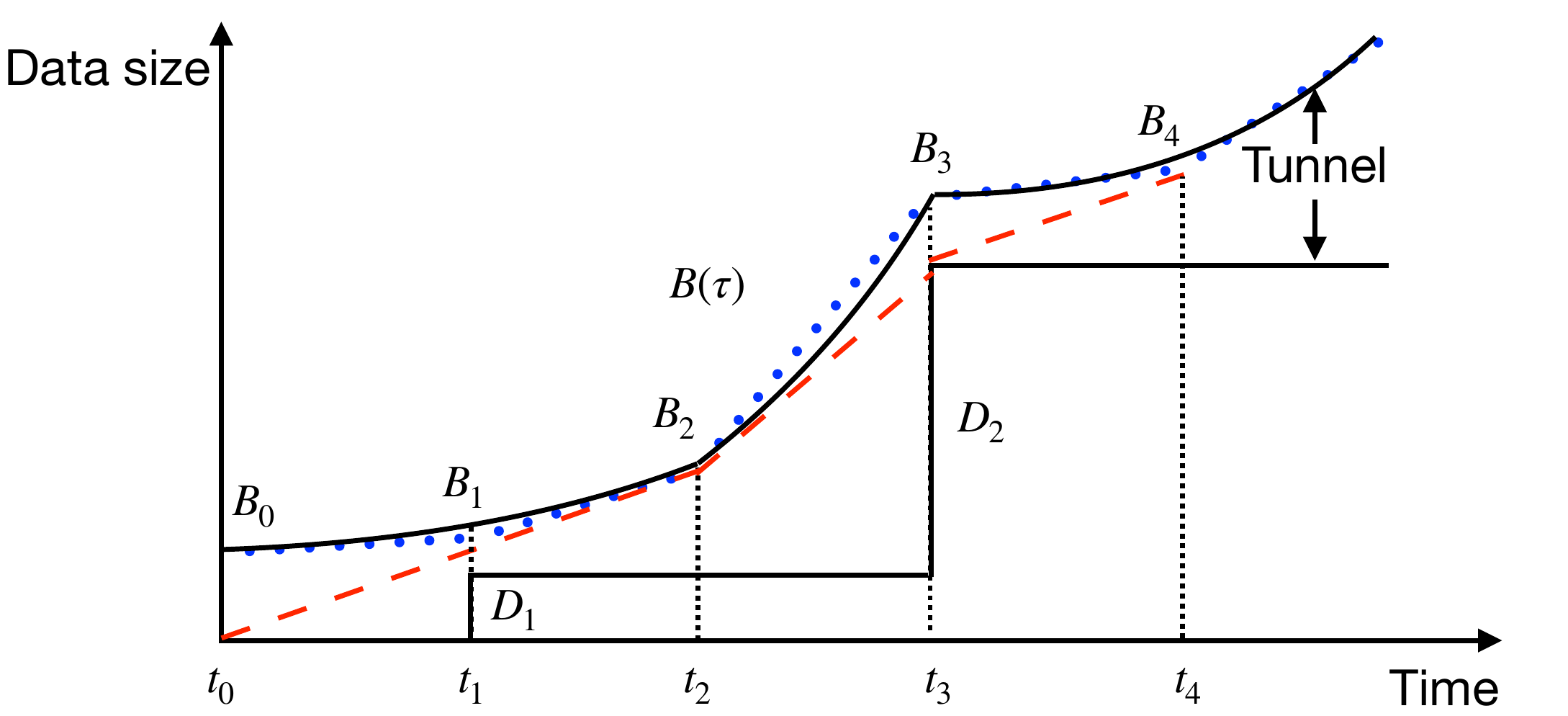}
\caption{The feasible data transmission tunnel with bursty data arrival.}
\label{FigBursty}
\end{figure}

In this section, a more practical scenario is considered where the data for model training is not stored in the sensor at the beginning but arrives during transmissions. As depicted by the upper solid curve in Fig.~\ref{FigBursty}, the amount of data collected by sensor is a continuous non-decreasing function with time denoted by $B(\tau)$, which adds a series of constraints as follows
\begin{equation}\label{Eq:Bursty}
\int_{t=0}^{\tau} r(t) dt \leq B(\tau).~\text{(data arrival constraints)}
\end{equation}
The resultant problem formulation is given as
\begin{subequations}
\begin{align}
\min_{\{D_n \geq 0\},\{r(t) \geq 0\}}  ~
& \bold{f}(\{D_n\},r(t)) \label{Eq:P6a}\\
\text{(P6)} \qquad \text{s.t.} \qquad
& \sum_{n=1}^j D_n \leq \int_{t=0}^{t_j} r(t) dt \leq B(t_j), j = 1,...N,\label{Eq:P6b}\\
& \int_{t=0}^{\tau} r(t) dt \leq B(\tau), \label{Eq:P6c}\\
& \sum_{n=1}^N D_n \leq D, \label{Eq:P6d}
\end{align}
\end{subequations}
which can be solved by the stratified sequencing or objectives merging as elaborated below.

\subsection{Stratified Sequencing for the Case of Bursty Data Arrival}
Problem (P6) can be divided into two sub-problems w.r.t. training error minimization and energy consumption minimization. The former sub-problem can be formulated as 

\begin{subequations}
\begin{align}
\min_{\{D_n \geq 0\}}  ~
& \sum_{n=1}^N \beta_n a_n \left(\frac{D_n}{d_n} + c_n\right)^{-b_n} \label{Eq:P6a1}\\
\text{(P6a)} \qquad \text{s.t.} \quad
& \sum_{n=1}^N D_n \leq D, \label{Eq:P6a2}\\
& \sum_{n=1}^j D_n \leq B(t_j), j = 1,...N.\label{Eq:P6a3}
\end{align}
\end{subequations}
It can be easily observed that (P6a) is a convex problem with convex objective function and linear constraints. Therefore, the optimal data partition $\{D_n^*\}$ can be obtained by applying the CVX ToolBox. As for energy consumption minimization, it should be noted that the continuous function $B(\tau)$ impedes the discretization of optimization variables $\{r(t)\}$. For the tractability concern, the data arrival curve is approximated by $M$ segments with total $\{B(t_m)\}$ bits of data collected by the instants $\{t_m\}$ as depicted by the dotted lines in Fig.~\ref{FigBursty}, where $\{t_n\} \subset \{t_m\}$ and $t_N = t_M$.\footnote{The continuous function $B(\tau)$ is recovered when $M \to \infty$.} The corresponding problem can be formulated as
\begin{subequations}
\begin{align}
\min_{\{r(t) \geq 0\}}  ~
& \int_{t=0}^{t_N} \left(e^{r(t)/B}-1\right)\frac{\sigma^2 T_n}{h} dt \label{Eq:P6b1}\\
\text{(P6b)} \qquad \text{s.t.} \quad
& \int_{t=0}^{t_j} r(t) dt \geq \sum_{n=1}^j D_n, j = 1,...N,\label{Eq:P6b2}\\
& \int_{t=0}^{t_m} r(t) dt  \leq B(t_m), m = 1,...M. \label{Eq:P6b3}
\end{align}
\end{subequations}

According to Lemma~\ref{Lemma:Constant}, the optimal transmission rate between any two successive instants of data arrival or task starting should be a constant. Following the similar proof of Lemma~\ref{Lemma:Condition} and \ref{Lemma:Direction}, one can observe that for the optimal transmission policy, the transmission rate decreases only at instants when the data buffer of the server is empty, and increases only at data arrival instant before which all the stored data at sensor is transmitted. Denote the subsequence of $\{t_m\}$ at which the transmission rate changes as $\{t_{v_i}\}$, the transmission rate has to be the form of
\begin{equation}
r(t) = r_i,~\forall t \in [t_{v_{i-1}}, t_{v_i}).
\end{equation}
It can be observed that the problem with bursty data arrival also has the structure of \emph{shifted} optimization problem as illustrated in the counterpart with limited data buffer capacity. Let $r_{\text{low}}[m]$ denote the lowest constant rate that can satisfy the data transmission constraint and $r_{\text{high}}[m]$ denote the highest constant rate that can satisfy the data arrival constraint in $[0,t_m]$. The range of constant-rate levels that would be feasible for the $m$-th epoch is defined by the interval $\bold{r}[m] = [r_{\text{low}}[m],r_{\text{high}}[m]]$. Therefore, we have
\begin{subequations}\label{Eq:intervalRand}
\begin{align}
& r_{\text{low}}[m] = \frac{\sum_{n: 0 \leq t_n < t_m} D_n}{t_m},\\
& r_{\text{high}}[m] = \frac{B(t_m)}{t_m},\\
& \bold{r}[m] = [r_{\text{low}}[m],r_{\text{high}}[m]] = \{r|r_{\text{low}}[m] \leq r \leq r_{\text{high}}[m]\},
\end{align}
\end{subequations}
for $m = 1,...,M$. The upper bound $v_b$ on the length of the first constant-rate transmission can be calculated as
\begin{equation}\label{Eq:boundRand}
v_b = \max\left\{v|\bigcap_{m=1}^u\bold{r}[m]\neq \emptyset, m=1,2,...,M\right\}.
\end{equation}
Given the sets of $\{r_e[j]\}$ and $\{r_f[j]\}$, assume that a constant rate $r_1$ over duration $t_{v_1}$ is feasible. This transmission then satisfies $r_1 \in \bigcap_{m=1}^{v_1}\bold{r}[m]$ and cannot extend beyond $t_{v_b}$, as it is rendered infeasible at $t_{v_b+1}$ by one of the constraints. A transmission with constant-rate $r_1$ either fails to satisfy the data transmission or arrival constraint at $t_{v_b+1}$. The former case implies that the rate after $t_{v_1}$ needs to increase, and the latter implies that the rate needs to decrease. Since an increase or decrease in rate can occur only at the instant when the data arrival or transmission constraint is active, the choice of $r_1$ in the optimal policy is restricted to $r_{\text{low}}[v_1]$ and $r_{\text{high}}[v_1]$ respectively for the two cases in consideration, where $v_1 = \max\left\{v|r_e[v] \in \bigcap_{m=1}^{v_b}\bold{r}[m]\right\}$ if $\bold{r}[v_b+1]$ falls below $\bigcap_{m=1}^{v_b}\bold{r}[m]$, otherwise $v_1 = \max\left\{v|r_f[v] \in \bigcap_{j=1}^{v_b}\bold{r}[m]\right\}$. Based on the above findings, the optimal transmission rate is determined by Algorithm~\ref{Al:OptBursty} following the similar proof of Theorem~\ref{Theorem:OptLimited}.


\begin{algorithm}[tt]
\renewcommand{\algorithmicrequire}{\textbf{Input:}}
\renewcommand{\algorithmicensure}{\textbf{Output:}}
\caption{String Pulling for Rate Control in Bursty Data Arrival Case.}
\label{Al:OptBursty}
\begin{algorithmic}[1]
\REQUIRE required data sizes $\{D_n\}$ at instants $\{t_n\}$, total data amount $\{B(t_m)\}$ at instants $\{t_m\}$.
\ENSURE the optimal transmission rates $\{r_i^*\}$ and durations $\{T_i^*\}$.
\STATE Initialize $v_b = 0$, $v_1 = 0$, $i = 0$, $n_1 = 0$.
\STATE \textbf{while} $M > 0$
\STATE \qquad Update $i = i+1$.
\STATE \qquad Update $r_{\text{low}}[m]$, $r_{\text{high}}[m]$, and $\bold{r}[m]$ for $m = 1,...,M$ according to \eqref{Eq:intervalRand}.
\STATE \qquad Update $v_b$ according to \eqref{Eq:boundRand}.
\STATE \qquad \textbf{if} $v_b = M$ 
\STATE \qquad \qquad Update $v_1 = \max\left\{v|r_{\text{low}}[v] \in \bigcap_{j=1}^{v_b}\bold{r}[m]\right\}$, $r_i^* = r_e[v_1]$, $T_i^* = t_{v_1}$.
\STATE \qquad \textbf{else} 
\STATE \qquad \qquad \textbf{if} $\bold{r}[v_b+1]$ falls below $\bigcap_{m=1}^{v_b}\bold{r}[m]$
\STATE \qquad \qquad \qquad Update $v_1 = \max\left\{v|r_{\text{low}}[v] \in \bigcap_{m=1}^{v_b}\bold{r}[m]\right\}$, $r_i^* = r_{\text{low}}[v_1]$ and $T_i^* = t_{v_1}$.
\STATE \qquad \qquad \textbf{else}
\STATE \qquad \qquad \qquad Update $v_1 = \max\left\{v|r_{\text{high}}[v] \in \bigcap_{m=1}^{v_b}\bold{r}[m]\right\}$, $r_i^* = r_{\text{high}}[v_1]$ and $T_i^* = t_{v_1}$.
\STATE \qquad \qquad \textbf{End if}
\STATE \qquad \textbf{End if}
\STATE \qquad Update $M \!=\! M \!-\! v_1$, $t_m \!=\! t_{m+v_1} \!-\! t_{v_1}$, $n_1 \!=\! \max\{n|t_n \!\leq\! t_{v_1}\}$, $t_n \!=\! t_{n+n_1} \!-\! t_{v_1}$, $t_0 \!=\! 0$.
\STATE \qquad Update $D' = r_i^* T_i^* - \sum_{n=0}^{n_1} D_n$, $D_n = D_{n+n_1}$, $D_1 = D_1 - D'$, $B(t_m) = B(t_{m+v_1}) - r_i^* T_i^*$.
\STATE \textbf{End while}
\STATE \textbf{Return} the optimal transmission rates $\{r_i^*\}$ and durations $\{T_i^*\}$.
\end{algorithmic}
\end{algorithm}

\subsection{Objectives Merging for the Case of Bursty Data Arrival}
By approximating the data arrival curve and applying weighted summation, the problem can be formulated as
\begin{subequations}
\begin{align}
\min_{\{D_n \geq 0\},\{r_m \geq 0\}}  ~
& \alpha \sum_{m=1}^{M} \left(e^{r_m/B}-1\right)\frac{\sigma^2 T_m}{h} + \sum_{n=1}^N \beta_n a_n \left(\frac{D_n}{d_n} + c_n\right)^{-b_n} \label{Eq:P7a}\\
\text{(P7)} \qquad \text{s.t.} \qquad
& \sum_{n:0 \leq t_n \leq t_k} D_n \leq \sum_{m=1}^k r_m T_m \leq B(t_k), k = 1,...M, \label{Eq:P7b}\\
& \sum_{n=1}^N D_n \leq D. \label{Eq:P7c}
\end{align}
\end{subequations}
It can be easily proved that problem (P7) is a convex problem with convex objective and linear constraints. Given the specific weights, the problem can be solved by the CVX Toolbox.

\section{Simulation Results}
This section provides simulation results to evaluate the performance of the proposed algorithms. Each point in the figures is obtained by averaging over 10 simulation runs, with independent channels in each run. All the algorithms are programed in Matlab R2021a on a desktop with Intel Xeon E5-1620 v3 CPU at 3.5 GHz and 32GB RAM. All the optimization problems after objectives merging are solved by the CVX Toolbox in the same Matlab platform. All the classifiers are trained by Python 3.6.12 on a GPU server with Intel Xeon E5-2678 v3 CPU at 2.5 GHz and NVIDIA TITAN XP GPU.


\subsection{Parameter Settings}
For the EI system, we consider the case of $N = 5$ different learning tasks at the server: 1) Classification of Scikit-learn dataset \cite{pedregosa2011scikit} via \emph{support vector machines} (SVM); 2) Classification of MNIST dataset \cite{deng2012mnist} via \emph{convolutional neural networks} (CNN); 3) Classification of Fashion-MNIST dataset \cite{xiao2017fashion} via CNN; 4) Classification of CIFAR-10 dataset \cite{he2016deep} via \emph{32-layer deep residual network} (ResNet-32); 5) Classification of 3D point clouds dataset ModelNet40 \cite{qi2017pointnet} via PointNet. These $5$ tasks are expected to be executed at instants $\{10000,20000,30000,50000,100000\}$s, respectively. The architectures of the neural networks are depicted in Fig.~\ref{FigSys}. 

In task-$1$, the SVM uses the penalty coefficient $C = 1$ and Gaussian kernel function $K(\bold{x}_i,\bold{x}_j) = \exp(-\gamma \times ||\bold{x}_i - \bold{x}_j||^2)$ with $\gamma = 0.001$. The SVM classifier is trained on the digit dataset in the Scikit-learn Python machine learning toolbox, which contains $1797$ images of size $8 \times 8$ from $10$ classes, with $5$ bits (corresponding to integers $0$ to $16$) for each pixel. Therefore, each data sample contains $D_n = 8 \times 8 \times 5 + 4 = 324$ bits ($4$ bits are reserved for the labels of $10$ classes). Out of all images, the SVM is trained using the first $1000$ samples and the later $797$ samples are used for testing. By varying the sample size $x_1$ as $\{x_1^{(o)}\} = \{30, 50, 70, 100, 300, 500, 700, 1000\}$, one can obtain the classification error $\Phi_1(x_1^{(o)})$ for each sample size $x_1^{(o)}$, where $o =1,...,O$ with $O = 8$ denoting the number of points to be fitted. By applying the following non-linear least squares fitting
\begin{equation}\label{Eq:Fitting}
\min_{\{a_n \geq 0\},\{b_n \geq 0\}}  \frac{1}{O} \sum_{o=1}^O |\Phi_n(x_n^{(o)}) - e_n(x_n, a_n, b_n)|^2,
\end{equation}
the tuning parameters for task-$1$ can be obtained as $(a_1, b_1) = (8.58, 0.86)$.

The task-$2$ is to train a 6-layer CNN based on the public MNIST dataset. Since the handwritten digits in the MNIST dataset are grayscale images with $28 \times 28$ pixels (each pixel has $8$ bits), in this case each data sample contains $8 \times 28 \times 28 + 4 = 6276$ bits. The input image is sequentially fed into a $5 \times 5$ convolution layer (with ReLu activation, $32$ channels, and SAME padding), a $2 \times 2$ max pooling layer, then another $5 \times 5$ convolution layer (with ReLu activation, $64$ channels, and SAME padding), a $2 \times 2$ max pooling layer, a fully connected layer with $128$ units (with ReLu activation), and a softmax output layer (with $10$ outputs). The training procedure is implemented via Adam optimizer with a learning rate of $10^{-4}$ and a mini-batch size of $100$. After training for $5000$ iterations, the model is testified on a validation dataset with $1000$ unseen samples to compute the classification error. By varying the sample size $x_2$ as $\{x_2^{(o)}\} = \{50, 100, 300, 500, 1000, 3000, 5000, 10000\}$, the tuning parameters for task-$2$ can be obtained as $(a_2, b_2) = (3.94, 0.53)$.

As for task-$3$, the CNN model in task-$2$ is trained based on the Fashion-MNIST dataset, which comprises $28 \times 28$ grayscale images of $70000$ fashion products from $10$ categories, with $7000$ images per category. The training set has $60000$ images and the test set has $10000$ images. By varying the sample size $x_3$ as $\{x_3^{(o)}\} = \{50, 100, 300, 500, 1000, 3000, 5000, 10000\}$, the tuning parameters for task-$3$ are obtained as $(a_3, b_3) = (3.89, 0.52)$.

\begin{figure}[t]
\centering
\includegraphics[scale=0.6]{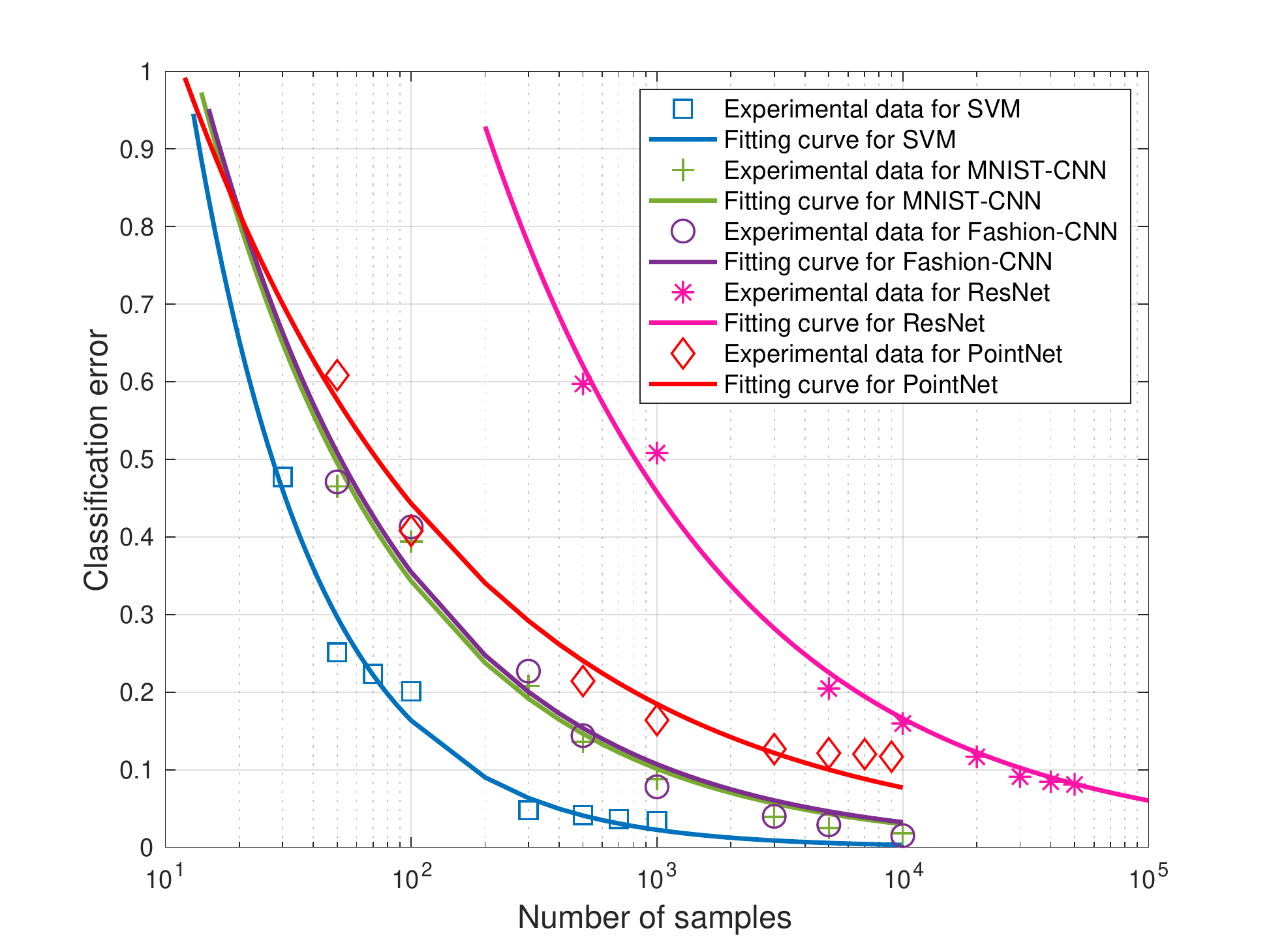}
\caption{Curve fitting for the learning tasks.}
\label{FigFitting}
\end{figure}

The task-$4$ is to train the ResNet-$32$ with $1.7$ M parameters using the CIFAR-$10$ dataset as the input images. The image in the CIFAR-$10$ dataset has $32 \times 32$ pixels (each pixel has $3$ Bytes representing RGB), and each image sample has a size of $(32 \times 32 \times 3 + 1) \times 8 = 24584$ bits. The training procedure is implemented with a diminishing learning rate and a mini-batch size of $100$. After training for $50000$ iterations, the trained model is tested on a dataset with $10000$ unseen samples, and obtain the corresponding classification error. By varying the sample size $x_4$ as $\{x_4^{(o)}\} = \{500, 1000, 5000, 10000, 20000, 30000, 40000, 50000\}$, the tuning parameters for task-$4$ are obtained as $(a_4, b_4) = (9.56, 0.44)$.

As for task-$5$, the PointNet with $3.5$ M parameters applies feature transformations and aggregates point features by max pooling to classify 3D point clouds dataset ModelNet-$40$. In ModelNet-$40$, there are $12311$ CAD models from $40$ object categories, split into $9843$ for training and $2468$ for testing. Each sample has $2000$ points with three single precision floating-point coordinates ($4$ Bytes), and the data size per sample is $(2000 \times 3 \times 4 + 1) \times 8 = 192008$ bits. Given the sample size $x_5$ as $\{x_5^{(o)}\} = \{50, 100, 500, 1000, 3000, 5000, 7000, 9000\}$, after training for $250$ epochs with a mini-batch size of $32$, the tuning parameters for task-$5$ are obtained as $(a_5, b_5) = (2.55, 0.38)$.

The fitting curves for the above $5$ tasks are depicted in Fig.~\ref{FigFitting}. The number of local samples at the server for the $5$ tasks are assumed to be $\{10,100,100,400,100\}$, respectively. The channels are under Rayleigh fading with bandwidth $B = 10^4$ Hz and noise power $\sigma^2 = 10^{-6}$ W. The weights of energy consumption and training errors are randomly selected with the summation equal to $1$. Despite our proposed joint design of data partition and rate control denoted by \emph{JDPRC}, $3$ benchmark schemes are considered for performance comparison. The \emph{EDP} scheme is to equally partition the data to $5$ tasks and optimize the rate control. The \emph{ERC} scheme is to optimize the data partition based on a constant transmission rate over the whole duration. The \emph{EDPRC} scheme adopts both the equal data partition and constant transmission rate.

\subsection{JDPRC for One-shot Data Arrival and Unlimited Data Buffer (ODAUDB) Case}
\begin{figure}[t]
\centering
\subfigure[]{
\label{FigCompare}
\includegraphics[scale=0.4]{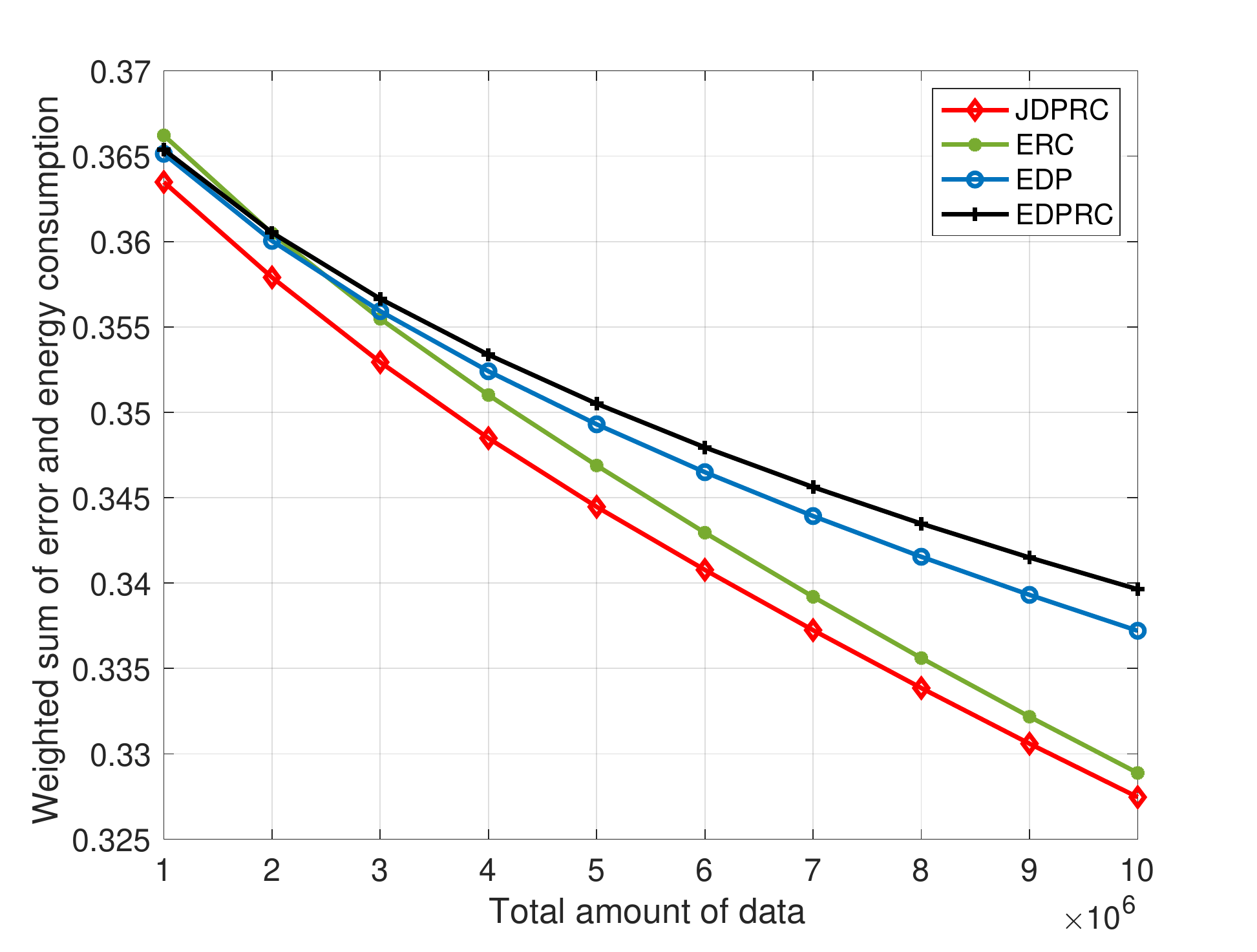}}
\subfigure[]{
\label{FigString}
\includegraphics[scale=0.4]{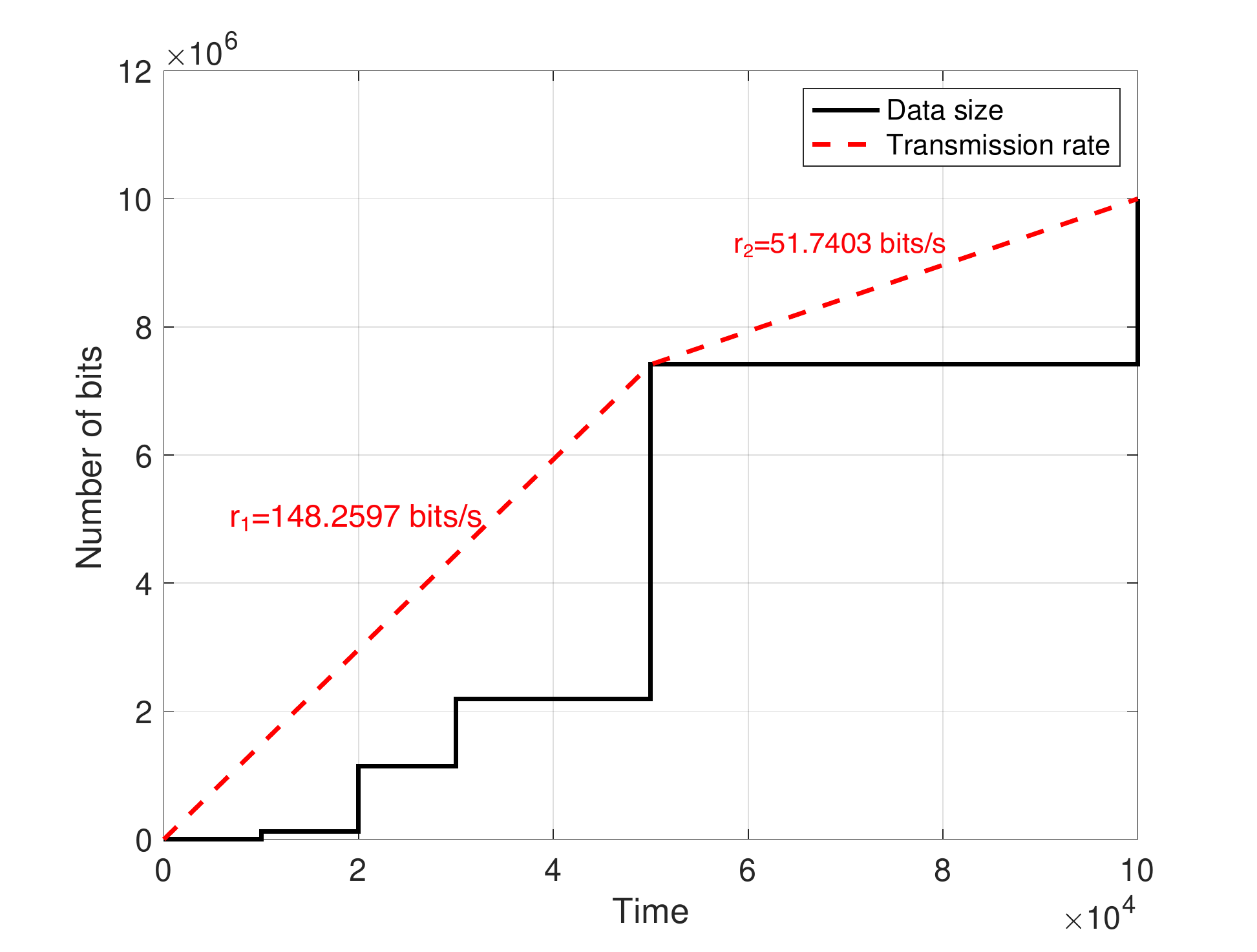}}
\subfigure[]{
\label{FigTest}
\includegraphics[scale=0.4]{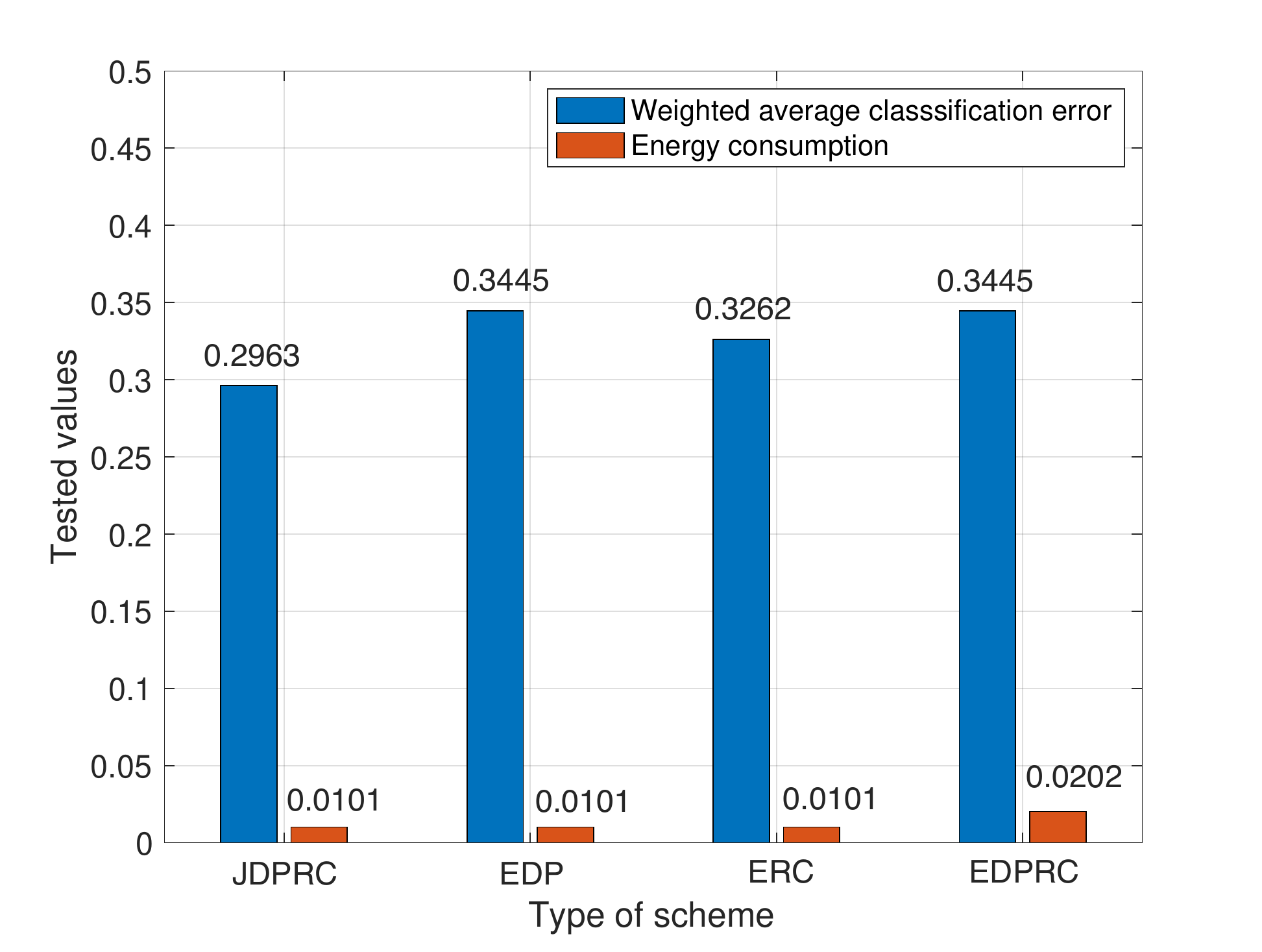}}
\subfigure[]{
\label{FigError}
\includegraphics[scale=0.4]{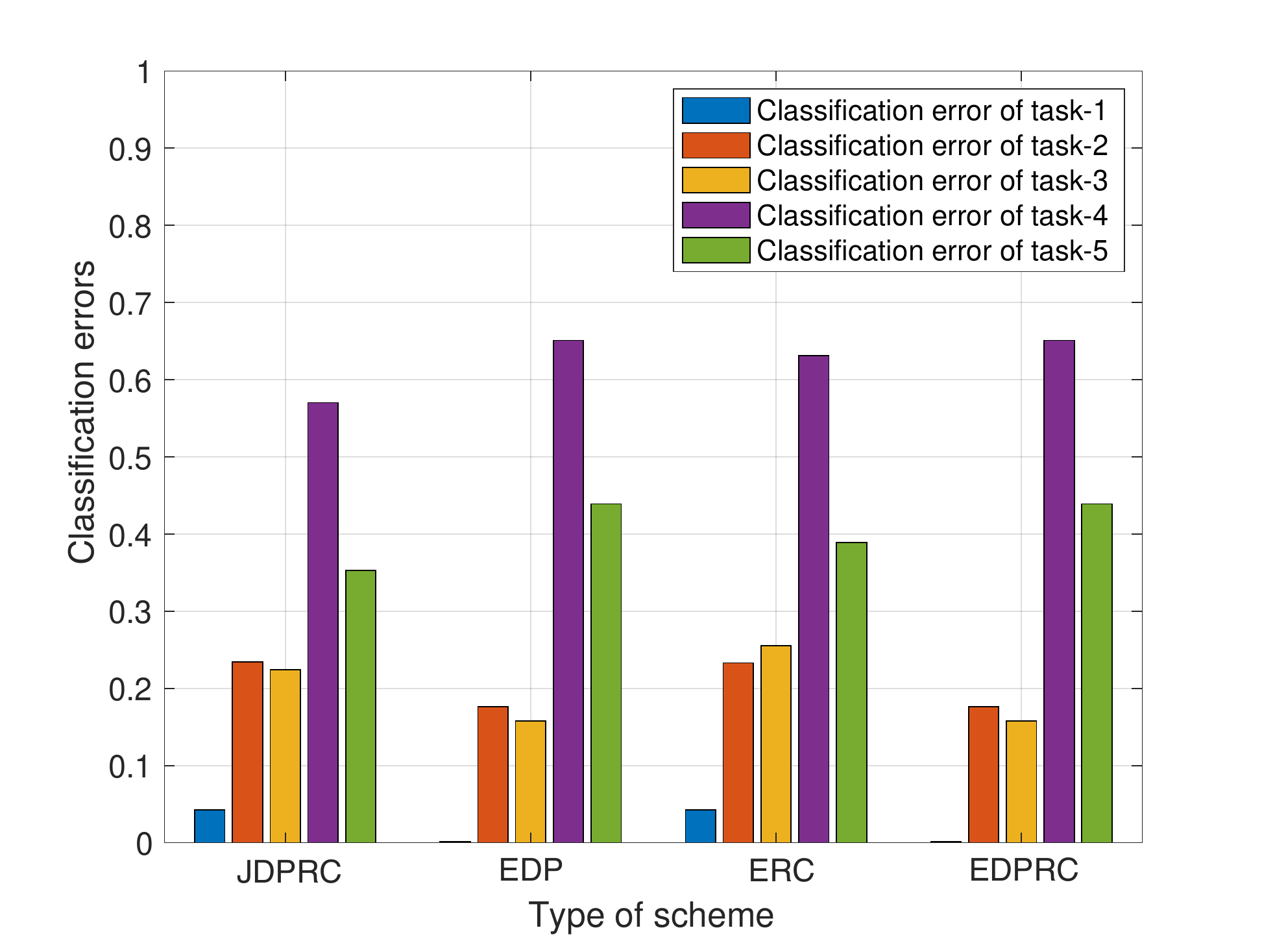}}
\caption{a) Performance metrics versus total amount of data with $N = 5$; b) Optimal structure of data partition and rate control with $N=5$ and $D = 10^7$ bits; c) Energy consumption and weighted average classification errors with $N=5$ and $D = 10^7$ bits; d) Classification errors of individual tasks with $N=5$ and $D = 10^7$ bits.}
\label{FigUnlimited}
\end{figure}

To examine our proposed JDPRC scheme, the performance metrics (i.e., weighted summation of energy consumption and testing errors) versus the total amount of data are plotted in Fig.~\ref{FigCompare}. It can be observed that the performance metrics monotonically decreases with the increasing the total amount of data, while the decreasing trend is becoming less steep. This is due to the fact that more available data can help reducing the classification errors to some extent, but such benefit disappears when there are enough data samples for model training. Moreover, our proposed JDPRC scheme can significantly reduces the weighted summation of energy consumption and classification errors compared with other $3$ benchmark schemes, which verifies the performance gain brought by the joint optimization. Given total $10^7$ bits of data samples, the specific data partition and rate control structures based on JDPRC scheme are depicted in Fig.~\ref{FigString}. One can observe that the optimal transmission rates are non-increasing and only decreases when the data buffer is empty, which is in accordance with the Lemma~\ref{Lemma:RateNon}.



To get deeper insights, given total $10^7$ bits of available data, the energy consumption and weighted average classification errors are illustrated in Fig.~\ref{FigTest}. It can be observed that our proposed JDPRC scheme can significantly reduce the testing error in general without exacerbating the energy consumption, which verifies the effectiveness of joint optimizing data partition and rate control. From Fig.~\ref{FigError}, it can be observed that the learning accuracies of the EDP, ERC, and EDPRC schemes are highly imbalanced due to excessive transmission of samples for learning tasks 1-2 and insufficient transmission of samples for learning tasks 3-5. The proposed JDPRC effectively mitigates such unfairness and achieves the desired balanced performance. Therefore, it achieves a significantly smaller weighted average classification error than that of other benchmark schemes.



\begin{figure}[t]
  \centering
  \subfigure[]{
  \label{FigComCase}
  \includegraphics[scale=0.4]{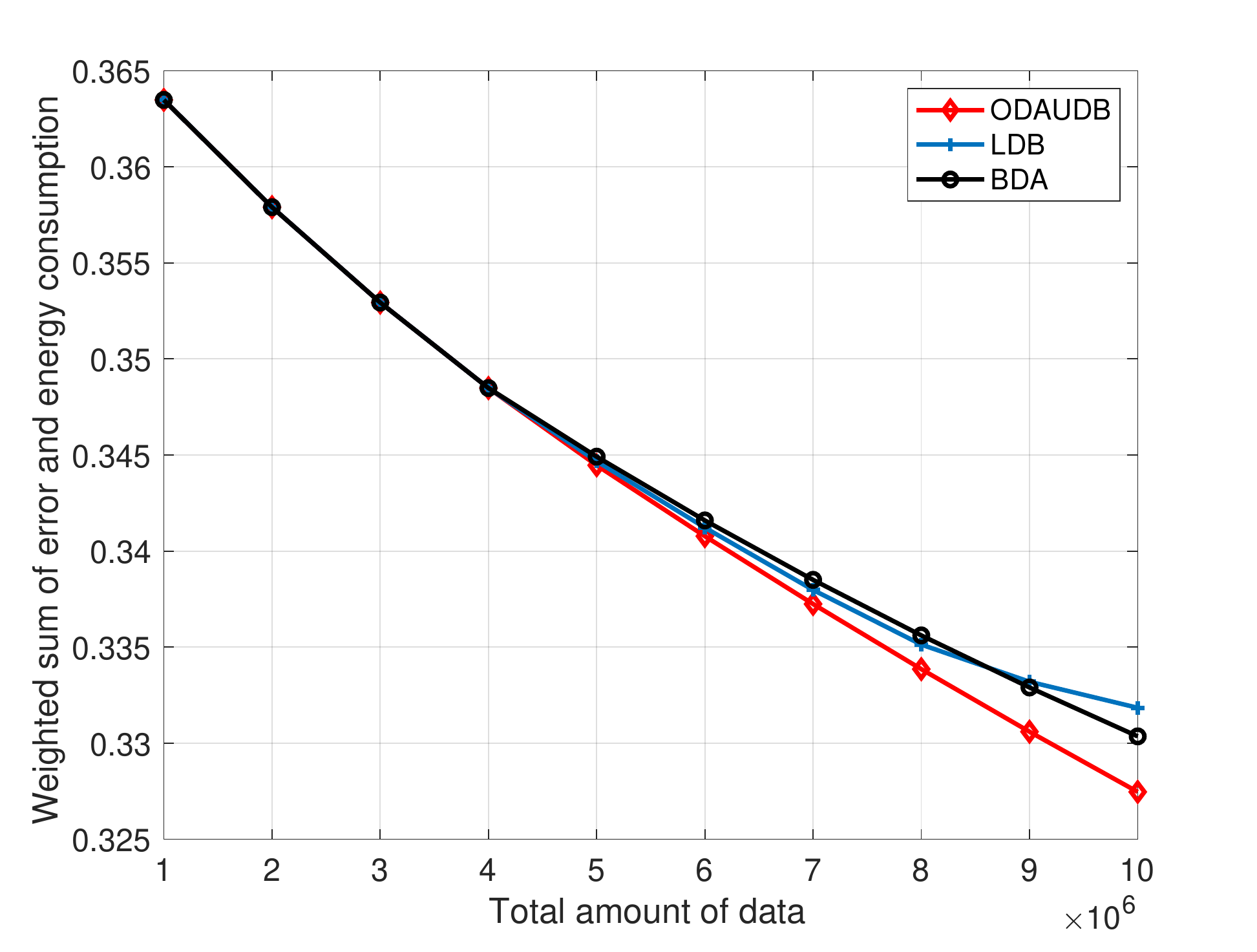}}
  \subfigure[]{
  \label{FigStrLimited}
  \includegraphics[scale=0.4]{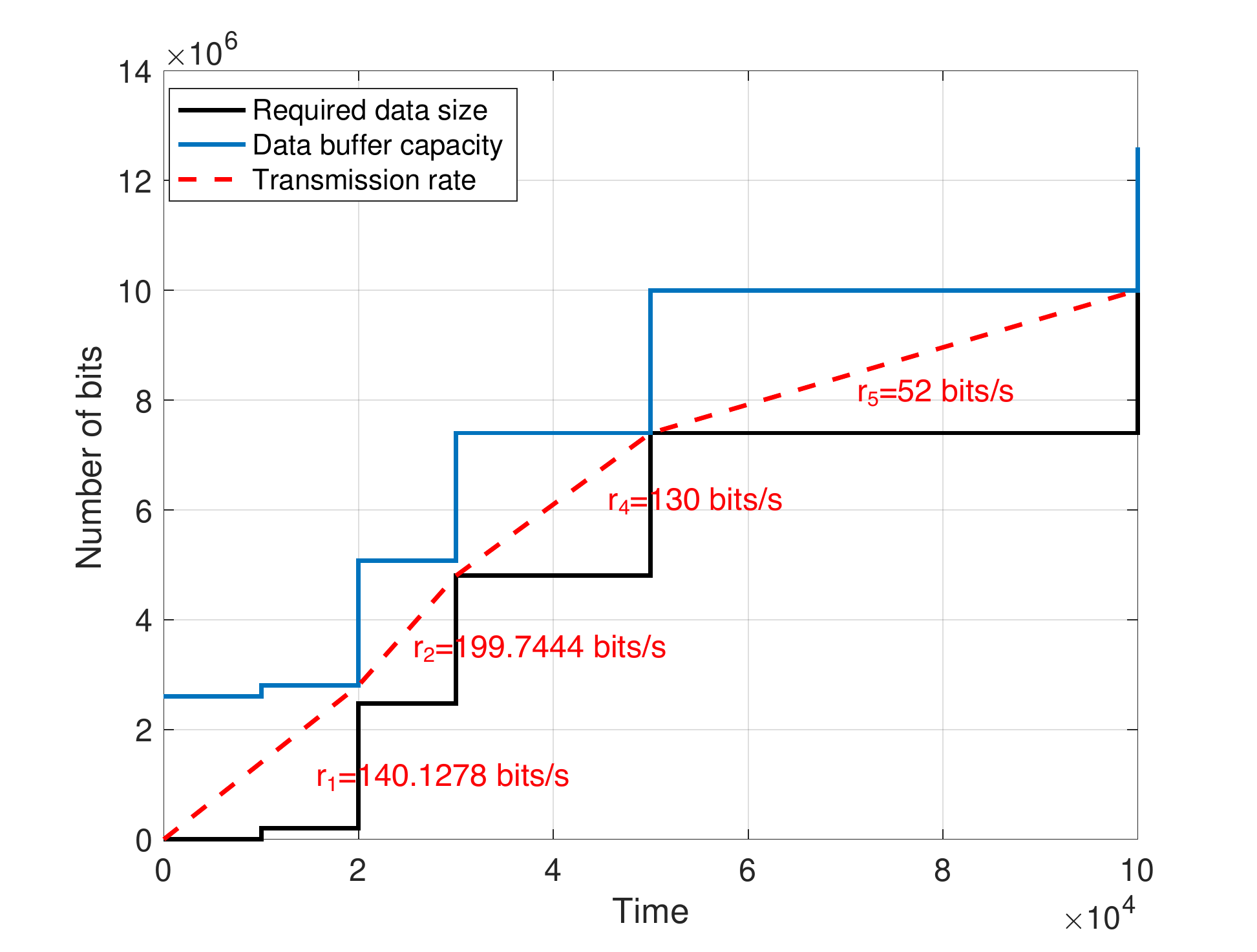}}
  \subfigure[]{
  \label{FigStrBursty}
  \includegraphics[scale=0.4]{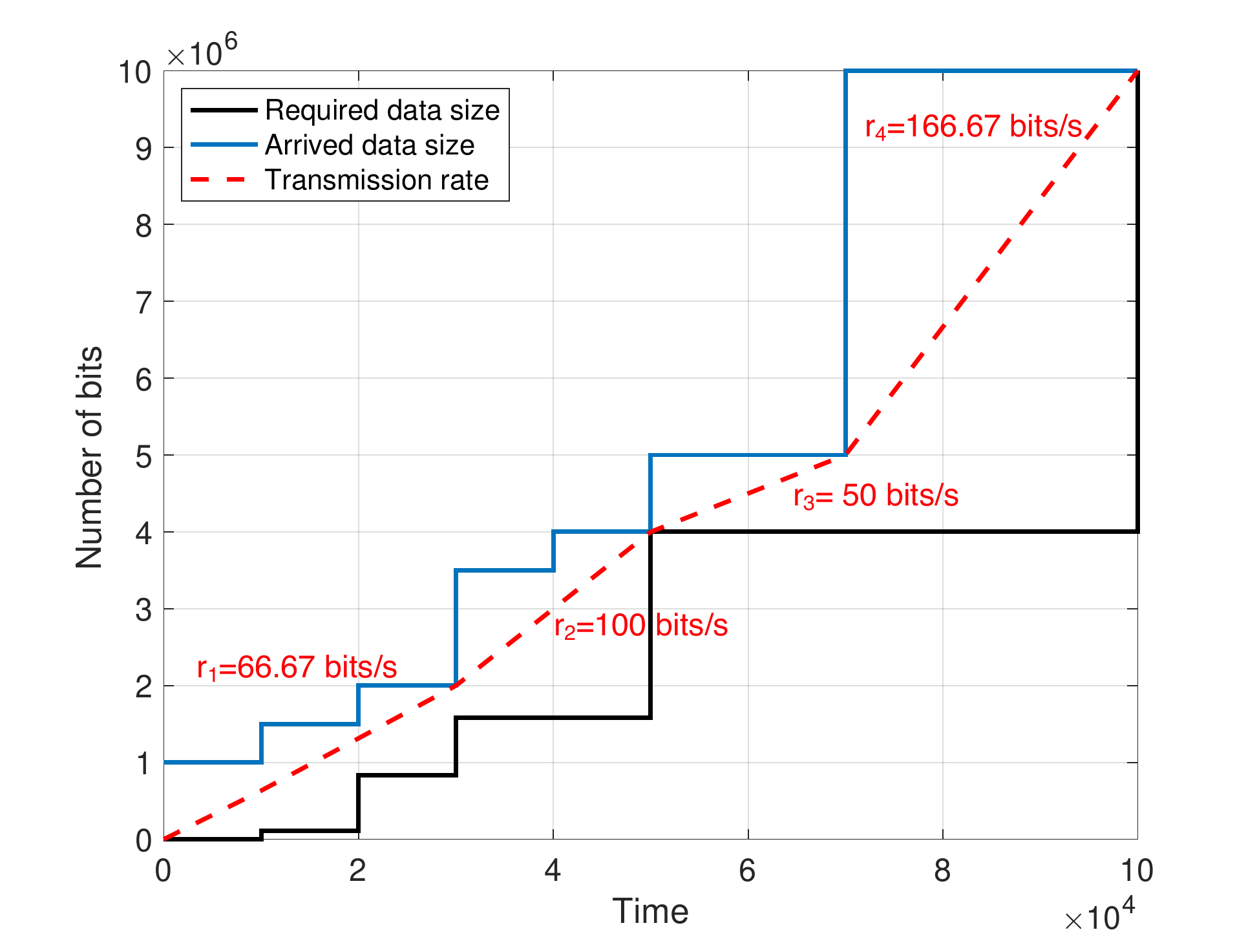}}
    \subfigure[]{
  \label{FigTestCase}
  \includegraphics[scale=0.4]{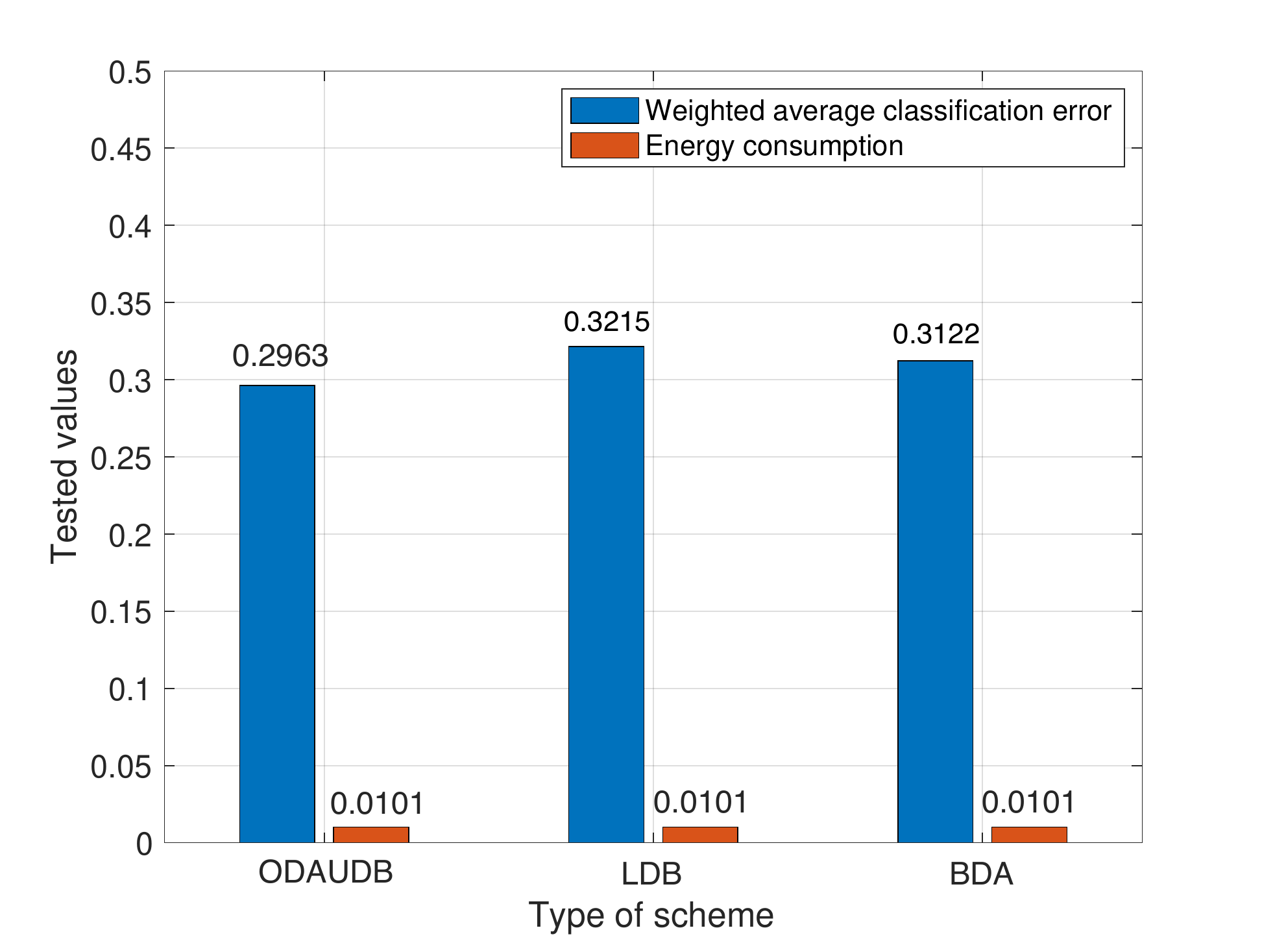}}
  \caption{a) Performance metrics versus total amount of data in different cases; b) Optimal structure of data partition and rate control in LDB case; c) Optimal structure of data partition and rate control in BDA case; d) Energy consumption and weighted average classification error in different cases.}
  \label{FigCase}
\end{figure}

\subsection{JDPRC for Limited Data Buffer (LDB) or Bursty Data Arrival (BDA) Cases}
The performance of our proposed JDPRC scheme is further testified in the cases with limited data buffer or bursty data arrival, respectively. Without loss of generality, the capacity of data buffer is set as $D_{\max} = 2.6 \times 10^6$ bits, while the arrived data amounts are approximately set as $[1, 1.5, 2, 3.5, 4, 5, 10] \times10^6$ bits at instants $\{1,2,3,4,5,7,10\} \times 10^4$ s.  As shown in Fig.~\ref{FigComCase}, the JDPRC scheme has similar performances in these three cases at the shortage of data, while the gap between them becomes larger with the increasing amount of data. This is due to the fact that the limited data buffer capacity in LDA case or limited amount of arrived data in BDA case becomes the main impediment for data partition given the abundant total amount of data. 

Given total $10^7$ bits of data samples, the specific data partition and rate control structures for LDB and BDA cases are depicted in Fig.~\ref{FigStrLimited} and Fig.~\ref{FigStrBursty}, respectively. One can observe that the transmission rate decreases only when the data buffer is empty and increases only when the data buffer is full in LDB case, which is in accordance with Lemma~\ref{Lemma:Condition} and \ref{Lemma:Direction}. As for BDA case, the transmission rate decreases only when the data buffer is empty, and increases only at data arrival instant before which all data at sensor is transmitted. As shown in Fig.~\ref{FigTestCase}, the energy consumptions caused by the JDPRC scheme are quite similar in all three cases, and the performance gap is mainly due to the difference between classification errors.

\section{Conclusion}
In this paper, we have investigated the design of data partition and rate control for learning-and-energy efficient long-term transmission. The joint design has been formulated as multi-objective optimization problems and solved by using convex optimization theory. In the case with finite server data buffer capacity and one-shot data arrival at sensor, the DWF structure is proved to be optimal for rate control and a SP algorithm is proposed to obtain the numerical values. Such findings are further extended to account for the cases with infinite server data buffer capacity or bursty data arrival at sensor. All the proposed schemes are testified on public datasets. This work opens a new direction for learning-and-energy efficient EI. The current design can be extended for more complex scenarios with multiple sensors and time-varying channels.

\appendix
\subsection{Proof of Lemma~\ref{Lemma:Constant}}\label{App:Constant}
Assume that there are two rates before and after instant $t_i \in [t_a,t_b)$, denoted as $r_i$ and $r_{i+1}$ respectively. The energy consumption is $E = \left(e^{r_i/B}-1\right)\frac{\sigma^2(t_i-t_a)}{h}+\left(e^{r_{i+1}/B}-1\right)\frac{\sigma^2(t_b-t_i)}{h}$. Let $r' = \frac{r_i (t_i - t_a) + r_{i+1} (t_b - t_i)}{t_b - t_a}$ as the new transmission rate over $[t_a,t_b)$, the transmit power becomes $P' = \left(e^{\frac{r_i (t_i - t_a) + r_{i+1} (t_b - t_i)}{(t_b - t_a)B}}-1\right)\frac{\sigma^2}{h}$. Due to the convexity, $P' \leq \frac{\left(e^{r_i/B}-1\right)\sigma^2}{h}\frac{t_i - t_a}{t_b - t_a} + \frac{\left(e^{r_{i+1}/B}-1\right)\sigma^2}{h}\frac{t_b - t_i}{t_b - t_a}$. The total energy consumption over this duration is $E' = \left(e^{\frac{r_i (t_i - t_a) + r_{i+1} (t_b - t_i)}{(t_b - t_a)B}}-1\right)\frac{\sigma^2(t_b - t_a)}{h} \leq E$. Therefore, the energy consumption under the new policy is less than that under the original policy for transmitting the same amount of data, thus the original policy cannot be optimal.

\subsection{Proof of Lemma~\ref{Lemma:RateNon}}\label{App:RateNon}
According to Eq.~\eqref{Eq:OptRate}, $r_{j+1}^* \leq r_j^*$ always holds. If any data for subsequent learning tasks is passed in epoch $j$, then the $j$-th constraint in Eq.~\eqref{Eq:P2b} is satisfied without equality. Therefore, $\mu_j = 0$ must holds according to the slackness conditions in Eq.~\eqref{Eq:KKTb2}. Hence, by Eq.~\eqref{Eq:OptRate}, $r_{j+1}^* = r_j^*$.

\subsection{Proof of Proposistion~\ref{Prop:OptRate}}\label{App:OptRate}
We will prove the necessity and sufficiency of the stated structure separately. First, we prove that the optimal policy must have the structure given above by contradiction. Assume the optimal policy satisfying Lemmas~\ref{Lemma:Constant} and \ref{Lemma:RateNon} does not has the structure above. Specifically, assume that the optimal policy over the duration $[0,t_{j_{i-1}}]$ is the same as the policy described in Proposition~\ref{Prop:OptRate}, while the transmission rate after $t_{j_{i-1}}$, which is $r_i$, is not the largest average rate starting from $t_{j_{i-1}}$. That is to say, we can find another $j' \leq N$, such that $r_i = \frac{\sum_{n=j_{i-1}+1}^{j_i} D_n}{t_{j_i}-t_{j_{i-1}}} < \frac{\sum_{n=j_{i-1}+1}^{j'} D_n}{t_{j'}-t_{j_{i-1}}} = r'$ (*). Based on Lemma~\ref{Lemma:RateNon}, the data transmitted up to $t_{j_{i-1}}$ equals to $\sum_{n=1}^{j_i-1} D_n$, i.e., there is no data remaining at $t = t_{j_{i-1}}^+$. We consider two possible cases here. The first case is $j' < j_i$. Under the optimal policy, the size of data that can be transmitted by rate $r_i$ over the duration $[t_{j_{i-1}},t_{j'}]$ is $r_i(t_{j'}-t_{j_{i-1}})$, which is smaller than the required data size $\sum_{n=j_{i-1}+1}^{j'} D_n$ and thus is infeasible. On the other hand, if $j' > j_i$, then the size of required data over $[t_{j_i},t_{j'}]$ is $\sum_{n=j_i+1}^{j'} D_n$. Since $\frac{\sum_{n=j_{i-1}+1}^{j'} D_n}{t_{j'}-t_{j_{i-1}}} = \frac{\sum_{n=j_{i-1}+1}^{j_i} D_n}{t_{j_i}-t_{j_{i-1}}} \frac{t_{j_i}-t_{j_{i-1}}}{t_{j'}-t_{j_{i-1}}} + \frac{\sum_{n=j_i+1}^{j'} D_n}{t_{j'}-t_{j_i}} \frac{t_{j'}-t_{j_i}}{t_{j'}-t_{j_{i-1}}}$, we have $r_i  < \frac{\sum_{n=j_{i-1}+1}^{j'} D_n}{t_{j'}-t_{j_{i-1}}} < \frac{\sum_{n=j_i+1}^{j'} D_n}{t_{j'}-t_{j_i}}$. Therefore, under any feasible policy, there must exists a duration $T \subseteq [t_{j_i},t_{j'}]$, such that the transmission rate over this duration is larger than $r_i$. This contradicts with Lemma~\ref{Lemma:RateNon} and thus the policy cannot be optimal. Next, we prove by contradiction that if a policy with rate vector $\mathbf{r}$ and duration vector $\mathbf{T}$ has the structure above, then it must be optimal. Assume there exists another policy with rate vector $\mathbf{r}'$ and $\mathbf{T}'$ such that the energy consumption $E'$ under this policy is smaller. We assume both of the policies are the same over the duration $[0,t_{j_{i-1}}]$, while the rate control policies after $t_{j_{i-1}}$ are different. The rates and durations are expressed by $r_i$, $r_i'$, $T_i$, and $T_i'$, respectively. Based on the assumption, we must have $r_i' < r_i$. From Lemma~\ref{Lemma:RateNon}, we know that the total required data over $(t_{j_{i-1}},t_{j_i}]$ should be equal to $r_i T_i$. If $T_i < T_i'$, since $r_i' T_i < r_i T_i$, $r_i'$ is infeasible. If $T_i > T_i'$, since $r_{i+1}' \leq r_i' < r_i$ according to Lemma~\ref{Lemma:RateNon}, one can find $r_i' T_i'+r_{i+1}'(T_i-T_i') < r_i T_i$, which means that $\mathbf{r}'$ is infeasible. Hence policy with $\mathbf{r}'$ cannot be optimal.

\subsection{Proof of Lemma~\ref{Lemma:Overflow}}\label{App:Overflow}
Due to the monotone increasing trend of transmitted data size, if data buffer overflows before any task starting instant $t_n$. Let the transmission rate allowing this overflow be $r(t)$. $D_n$ is less than or equal to $D_{\max}$ by system model, and thus the size of data in buffer at $t_n$ is strictly positive. This implies that $r(t)$ can be decreased by an infinitesimal amount $\delta$ in $(t_n-\epsilon, t_n)$ without violating data transmission constraints, which strictly reduces the energy consumption. Therefore, a transmission rate that yields a data buffer overflow can not be optimal.

\subsection{Proof of Lemma~\ref{Lemma:Condition}}\label{App:Condition}
Assume that the transmission rate changes at arbitrary time $t$, so that $r(t^-) \neq r(t^+)$. Consider the interval $[t-\tau,t+\tau]$, where the policy transmits a total data of $\tau(r(t^-)+r(t^+))$. Unless the data buffer is full or empty at $t$, the data feasibility constraints will be inactive in this interval. Let $r^*(t)=\frac{r(t^-)+r(t^+)}{2}$ be the constant rate in $[t-\tau,t+\tau]$ that transmits the same amount of data, which is feasible for a sufficiently small $\tau$. Then $r(t)$ can be replaced by $r^*(t)$ without alerting the rest of the schedule. According to Lemma~\ref{Lemma:Constant}, the new rate policy consumes strictly less energy. Therefore, $r(t)$ must stay constant unless either the data buffer is full or empty.

\subsection{Proof of Lemma~\ref{Lemma:Direction}}\label{App:Direction}
The proof is by contradiction. Consider the notation in the proof of Lemma~\ref{Lemma:Condition}. Non-empty data buffer at time $t$ implies that the data transmission constraint is not active. Therefore, if $r(t^-) > r(t^+)$ holds, i.e., if the transmission rate is decreasing at $t$, replacing $r(t^-)$ with $r^*(t)$ on $[t-\tau,t]$ slightly decreases $\int_{t^-}^{t} r(t) dt$, which is feasible. Similarly, data buffer not being full at time $t$ implies that the data buffer constraint is not active. Therefore, if $r(t^-) < r(t^+)$ holds, i.e., if the transmission rate is increasing at $t$, replacing $r(t^-)$ with $r^*(t)$ on $[t-\tau,t]$ slightly increases $\int_{t^-}^{t} r(t) dt$, which is feasible. Due to Lemma~\ref{Lemma:Constant}, the new policy $r^*(t)$ consumes strictly less energy.

\subsection{Proof of Theorem~\ref{Theorem:OptLimited}}\label{App:OptLimited}
If $\bold{r}[u_b+1]$ lies below $\bigcap_{j=1}^{u_b}\bold{r}[j]$, Algorithm~\ref{Al:OptLimited} suggests that a transmission of duration $t_{u_1}$ with rate $r_e[u_1]$ is optimal. This can be shown by contradiction. According to Lemma~\ref{Lemma:Constant}, a constant transmission rate is optimal in a time interval if it is feasible in that interval. Therefore, the optimal transmission rate must be a constant in $[0,t_{u_1}]$. Suppose there exist another constant rate $r' \neq r_e[u_1]$. If $r' < r_e[u_1]$, we have $r' t_{u_1} < r_e[u_1] t_{u_1} = \sum_{n=0}^{u_1} D_n$, which means that $r'$ fails to satisfy the data transmission constraint at $t_{u_1}$. If $r' > r_e[u_1]$, we have $r' t_{u_1} > r_e[u_1] t_{u_1} = \sum_{n=0}^{u_1} D_n$, which means that the data buffer is non-empty at $t_{u_1}$ and thus rate $r'$ cannot change according to Lemma~\ref{Lemma:Condition}. Since $\bold{r}[u_b+1]$ lies below $\bigcap_{j=1}^{u_b}\bold{r}[j]$, we have $\sum_{j=0}^{u_b} D_j +D_{\max} < r_e[u_b] t_{u_b+1} < r' t_{u_b+1}$, which means that $r'$ fails to satisfy the data buffer constraint at $t_{u_b+1}$. If $\bold{r}[u_b+1]$ lies above $\bigcap_{j=0}^{u_b}\bold{r}[j]$, Algorithm~\ref{Al:OptLimited} suggests that a transmission of duration $t_{u_1}$ with rate $r_f[u_1]$ is optimal. This can also be shown by contradiction. On one hand, the constant rate larger than $r_f[u_1]$ fails to satisfy the data buffer constraint at $t_{u_1}$. On the other hand, the constant rate less than $r_f[u_1]$ fails to satisfy the data transmission constraint at $t_{u_b+1}$.

\bibliographystyle{ieeetr}

\begin{thebibliography}{10}
\providecommand{\url}[1]{#1}
\csname url@samestyle\endcsname
\providecommand{\newblock}{\relax}
\providecommand{\bibinfo}[2]{#2}
\providecommand{\BIBentrySTDinterwordspacing}{\spaceskip=0pt\relax}
\providecommand{\BIBentryALTinterwordstretchfactor}{4}
\providecommand{\BIBentryALTinterwordspacing}{\spaceskip=\fontdimen2\font plus
\BIBentryALTinterwordstretchfactor\fontdimen3\font minus
  \fontdimen4\font\relax}
\providecommand{\BIBforeignlanguage}[2]{{%
\expandafter\ifx\csname l@#1\endcsname\relax
\typeout{** WARNING: IEEEtran.bst: No hyphenation pattern has been}%
\typeout{** loaded for the language `#1'. Using the pattern for}%
\typeout{** the default language instead.}%
\else
\language=\csname l@#1\endcsname
\fi
#2}}
\providecommand{\BIBdecl}{\relax}
\BIBdecl

\bibitem{mitchell2013artificial}
R.~Mitchell, J.~Michalski, and T.~Carbonell, \emph{An artificial intelligence
  approach}.\hskip 1em plus 0.5em minus 0.4em\relax Springer, 2013.

\bibitem{zhou2019edge}
Z.~Zhou, X.~Chen, E.~Li, L.~Zeng, K.~Luo, and J.~Zhang, ``Edge intelligence:
  Paving the last mile of artificial intelligence with edge computing,''
  \emph{Proc. IEEE}, vol. 107, no.~8, pp. 1738--1762, 2019.

\bibitem{yu2020intelligent}
S.~Yu, X.~Chen, L.~Yang, D.~Wu, M.~Bennis, and J.~Zhang, ``Intelligent edge:
  Leveraging deep imitation learning for mobile edge computation offloading,''
  \emph{IEEE Wireless Commun.}, vol.~27, no.~1, pp. 92--99, 2020.

\bibitem{mao2017survey}
Y.~Mao, C.~You, J.~Zhang, K.~Huang, and K.~B. Letaief, ``A survey on mobile
  edge computing: The communication perspective,'' \emph{IEEE Commun. Surveys
  Tuts.}, vol.~19, no.~4, pp. 2322--2358, 2017.

\bibitem{zhu2020toward}
G.~Zhu, D.~Liu, Y.~Du, C.~You, J.~Zhang, and K.~Huang, ``Toward an intelligent
  edge: Wireless communication meets machine learning,'' \emph{IEEE Commun.
  Mag.}, vol.~58, no.~1, pp. 19--25, 2020.

\bibitem{wang2018edge}
S.~Wang, T.~Tuor, T.~Salonidis, K.~K. Leung, C.~Makaya, T.~He, and K.~Chan,
  ``When edge meets learning: Adaptive control for resource-constrained
  distributed machine learning,'' in \emph{Proc. IEEE Infocom}, 2018, pp.
  63--71.

\bibitem{zhu2020one}
G.~Zhu, Y.~Du, D.~G{\"u}nd{\"u}z, and K.~Huang, ``One-bit over-the-air
  aggregation for communication-efficient federated edge learning: Design and
  convergence analysis,'' \emph{IEEE Trans. Wireless Commun.}, vol.~20, no.~3,
  pp. 2120--2135, 2020.

\bibitem{yang2020federated}
K.~Yang, T.~Jiang, Y.~Shi, and Z.~Ding, ``Federated learning via over-the-air
  computation,'' \emph{IEEE Transactions on Wireless Communications}, vol.~19,
  no.~3, pp. 2022--2035, 2020.

\bibitem{amiri2020machine}
M.~M. Amiri and D.~G{\"u}nd{\"u}z, ``Machine learning at the wireless edge:
  Distributed stochastic gradient descent over-the-air,'' \emph{IEEE
  Transactions on Signal Processing}, vol.~68, pp. 2155--2169, 2020.

\bibitem{zhang2021gradient}
N.~Zhang and M.~Tao, ``Gradient statistics aware power control for over-the-air
  federated learning,'' \emph{IEEE Transactions on Wireless Communications},
  vol. early access, 2021.

\bibitem{liu2020privacy}
D.~Liu and O.~Simeone, ``Privacy for free: {Wireless} federated learning via
  uncoded transmission with adaptive power control,'' \emph{IEEE J. Sel. Areas
  Commun.}, vol.~39, no.~1, pp. 170--185, 2020.

\bibitem{wang2020machine}
S.~Wang, Y.-C. Wu, M.~Xia, R.~Wang, and H.~V. Poor, ``Machine intelligence at
  the edge with learning centric power allocation,'' \emph{IEEE Trans. Wireless
  Commun.}, vol.~19, no.~11, pp. 7293--7308, 2020.

\bibitem{ren2020accelerating}
J.~Ren, G.~Yu, and G.~Ding, ``Accelerating {DNN} training in wireless federated
  edge learning systems,'' \emph{IEEE J. Sel. Areas Commun.}, vol.~39, no.~1,
  pp. 219--232, 2020.

\bibitem{wen2020joint}
D.~Wen, M.~Bennis, and K.~Huang, ``Joint parameter-and-bandwidth allocation for
  improving the efficiency of partitioned edge learning,'' \emph{IEEE Trans.
  Wireless Commun.}, vol.~19, no.~12, pp. 8272--8286, 2020.

\bibitem{chen2020joint}
M.~Chen, Z.~Yang, W.~Saad, C.~Yin, H.~V. Poor, and S.~Cui, ``A joint learning
  and communications framework for federated learning over wireless networks,''
  \emph{IEEE Trans. Wireless Commun.}, vol.~20, no.~1, pp. 269--283, 2021.

\bibitem{shi2020joint}
W.~{Shi}, S.~{Zhou}, Z.~{Niu}, M.~{Jiang}, and L.~{Geng}, ``Joint device
  scheduling and resource allocation for latency constrained wireless federated
  learning,'' \emph{IEEE Trans. Wireless Commun.}, vol.~20, no.~1, pp.
  453--467, 2021.

\bibitem{schwartz2020green}
R.~Schwartz, J.~Dodge, N.~A. Smith, and O.~Etzioni, ``Green {AI},''
  \emph{Commun. ACM}, vol.~63, no.~12, pp. 54--63, 2020.

\bibitem{guler2021sustainable}
B.~Guler and A.~Yener, ``Sustainable federated learning,'' \emph{[Online].
  Available: https://arxiv.org/pdf/2102.11274.pdf}, 2021.

\bibitem{guler2021energy}
------, ``Energy-harvesting distributed machine learning,'' \emph{[Online].
  Available: https://arxiv.org/pdf/2102.05639.pdf}, 2021.

\bibitem{zeng2021wirelessly}
Q.~Zeng, Y.~Du, and K.~Huang, ``Wirelessly powered federated edge learning:
  Optimal tradeoffs between convergence and power transfer,'' \emph{IEEE Trans.
  Wireless Commun.}, early access, 2021.

\bibitem{sun2020energy}
Y.~Sun, S.~Zhou, and D.~G{\"u}nd{\"u}z, ``Energy-aware analog aggregation for
  federated learning with redundant data,'' in \emph{Proc. IEEE ICC}, Dublin,
  Ireland, July 2020.

\bibitem{yang2020energy}
Z.~Yang, M.~Chen, W.~Saad, C.~S. Hong, and M.~Shikh-Bahaei, ``Energy efficient
  federated learning over wireless communication networks,'' \emph{IEEE
  Transactions on Wireless Communications}, vol.~20, no.~3, pp. 1935--1949,
  2021.

\bibitem{mo2020energy}
X.~Mo and J.~Xu, ``Energy-efficient federated edge learning with joint
  communication and computation design,'' \emph{[Online]. Available:
  https://arxiv.org/pdf/2003.00199.pdf}, 2020.

\bibitem{zeng2020energy}
Q.~Zeng, Y.~Du, K.~Huang, and K.~K. Leung, ``Energy-efficient resource
  management for federated edge learning with {CPU-GPU} heterogeneous
  computing,'' \emph{IEEE Trans. Wireless Commun.}, early access, 2021.

\bibitem{prabhakar2001energy}
B.~Prabhakar, E.~U. Biyikoglu, and A.~El~Gamal, ``Energy-efficient transmission
  over a wireless link via lazy packet scheduling,'' in \emph{Proc. IEEE
  Infocom}, Anchorage, AK, USA, Aug. 2001.

\bibitem{zafer2005calculus}
M.~A. Zafer and E.~Modiano, ``A calculus approach to minimum energy
  transmission policies with quality of service guarantees,'' in \emph{Proc.
  IEEE Infocom}, Miami, FL, USA, Aug. 2005.

\bibitem{zafer2009calculus}
------, ``A calculus approach to energy-efficient data transmission with
  quality-of-service constraints,'' \emph{IEEE/ACM Trans. Netw.}, vol.~17,
  no.~3, pp. 898--911, 2009.

\bibitem{yang2011optimal}
J.~Yang and S.~Ulukus, ``Optimal packet scheduling in an energy harvesting
  communication system,'' \emph{IEEE Trans. Commun.}, vol.~60, no.~1, pp.
  220--230, 2011.

\bibitem{tutuncuoglu2012optimum}
K.~Tutuncuoglu and A.~Yener, ``Optimum transmission policies for battery
  limited energy harvesting nodes,'' \emph{IEEE Trans. Wireless Commun.},
  vol.~11, no.~3, pp. 1180--1189, 2012.

\bibitem{you2018exploiting}
C.~You and K.~Huang, ``Exploiting non-causal {CPU}-state information for
  energy-efficient mobile cooperative computing,'' \emph{IEEE Trans. Wireless
  Commun.}, vol.~17, no.~6, pp. 4104--4117, 2018.

\bibitem{domhan2015speeding}
T.~Domhan, J.~T. Springenberg, and F.~Hutter, ``Speeding up automatic
  hyperparameter optimization of deep neural networks by extrapolation of
  learning curves,'' in \emph{Proc. AAAI conference on artificial
  intelligence}, Austin Texas, USA, Jan. 2015.

\bibitem{johnson2018predicting}
M.~Johnson, P.~Anderson, M.~Dras, and M.~Steedman, ``Predicting accuracy on
  large datasets from smaller pilot data,'' in \emph{Proc. Annual Meeting of
  ACL}, Melbourne, Australia, July 2018.

\bibitem{liu2020edge}
D.~Liu, S.~Wang, Z.~Wen, L.~Cheng, M.~Wen, and Y.-C. Wu, ``Edge learning with
  unmanned ground vehicle: Joint path, energy, and sample size planning,''
  \emph{IEEE Internet Things J.}, vol.~8, no.~4, pp. 2959--2975, 2020.

\bibitem{zhou2020learning}
L.~Zhou, Y.~Hong, S.~Wang, R.~Han, D.~Li, R.~Wang, and Q.~Hao, ``Learning
  centric wireless resource allocation for edge computing: Algorithm and
  experiment,'' \emph{IEEE Trans. Veh. Technol.}, vol.~70, no.~1, pp.
  1035--1040, 2020.

\bibitem{seung1992statistical}
H.~S. Seung, H.~Sompolinsky, and N.~Tishby, ``Statistical mechanics of learning
  from examples,'' \emph{Phys. Rev. A}, vol.~45, no.~8, pp. 6056--6091, 1992.

\bibitem{ozel2011transmission}
O.~Ozel, K.~Tutuncuoglu, J.~Yang, S.~Ulukus, and A.~Yener, ``Transmission with
  energy harvesting nodes in fading wireless channels: Optimal policies,''
  \emph{IEEE J. Sel. Areas Commun.}, vol.~29, no.~8, pp. 1732--1743, 2011.

\bibitem{grant2009cvx}
M.~Grant, S.~Boyd, and Y.~Ye, ``{CVX}: Matlab software for disciplined convex
  programming,'' \emph{[Online]. Available: http://stanford.edu/boyd/cvx},
  2009.

\bibitem{pedregosa2011scikit}
F.~Pedregosa, G.~Varoquaux, A.~Gramfort \emph{et~al.}, ``Scikit-learn: Machine
  learning in python,'' \emph{J. Mach. Learn. Res.}, vol.~12, pp. 2825--2830,
  2011.

\bibitem{deng2012mnist}
L.~Deng, ``The {MNIST} database of handwritten digit images for machine
  learning research,'' \emph{IEEE Signal Process. Mag.}, vol.~29, no.~6, pp.
  141--142, 2012.

\bibitem{xiao2017fashion}
H.~Xiao, K.~Rasul, and R.~Vollgraf, ``Fashion-{MNIST}: a novel image dataset
  for benchmarking machine learning algorithms,'' \emph{[Online]. Available:
  https://arxiv.org/pdf/2102.11274.pdf}, 2017.

\bibitem{he2016deep}
K.~He, X.~Zhang, S.~Ren, and J.~Sun, ``Deep residual learning for image
  recognition,'' in \emph{Proc. CVPR}, Las Vegas, Nevada, US, June 2016.

\bibitem{qi2017pointnet}
C.~R. Qi, H.~Su, K.~Mo, and L.~J. Guibas, ``{PointNet}: Deep learning on point
  sets for {3D} classification and segmentation,'' in \emph{Proc. CVPR},
  Honolulu, Hawaii, US, July 2017.

\end{thebibliography}

\end{document}